
\def\la{\mathrel{\hbox{\rlap{\hbox{\lower4pt\hbox{$\sim$}}}\hbox{$<$}}}}
\def\ga{\mathrel{\hbox{\rlap{\hbox{\lower4pt\hbox{$\sim$}}}\hbox{$>$}}}}

\def\farcs{\hbox{$\> .\!\!^{\prime\prime}$}}
\def\fss{\hbox{$\> .\!\!^{\rm s}$}}

\def\vu{{\rm km\>s^{-1}}}
\def\msol{{M_{\odot}}}
\def\dH0{{H$_0=$100h \vu \>\rm Mpc$^{-1}\>$}}
\def\etal{{\it et~al.\/}}

\def\refpar{\par\hangindent=3em\hangafter=1}
\def\reference{\relax\refpar}

\font\smcap=cmcsc10

\magnification=\magstep1

\hsize 7.2truein
\hoffset -0.3truein

\voffset -0.25truein
\vsize 9.5truein
\overfullrule=0pt
\centerline{\bf GLOBULAR CLUSTER PHOTOMETRY WITH}
\centerline{{\bf THE HUBBLE SPACE TELESCOPE}\footnote{$^1$}{Based on
observations with the NASA/ESA {\it Hubble Space Telescope\/} obtained at the
Space Telescope Science Institute, which is operated by the Association of
Universities for Research in Astronomy, Inc., under NASA contract
NAS5-26555}{\bf . V.}}
\centerline{\bf WFPC2 STUDY OF M15'S CENTRAL DENSITY CUSP}
\vskip 0.7truein

\centerline{\bf Puragra Guhathakurta}
\centerline{UCO/Lick Observatory, University of California, Santa Cruz, CA
95064}
\centerline{Electronic mail: \tt raja@lick.ucsc.edu}

\bigskip
\centerline{\bf Brian Yanny}
\centerline{Fermi National Accelerator Laboratory, Batavia, IL 60510}
\centerline{Electronic mail: \tt yanny@sdss.fnal.gov}

\bigskip
\centerline{\bf Donald P.~Schneider}
\centerline{Department of Astronomy and Astrophysics}
\centerline{The Pennsylvania State University, University Park, PA 16802}
\centerline{Electronic mail: \tt dps@astro.psu.edu}

\medskip
\centerline{\bf and}

\medskip
\centerline{\bf John N.~Bahcall}
\centerline{Institute for Advanced Study, Princeton, NJ 08540}
\centerline{Electronic mail: \tt jnb@sns.ias.edu}

\vfill\eject

\centerline {ABSTRACT}
\medskip

We describe images of the center of the dense globular cluster M15 (NGC~7078)
obtained with the {\it Hubble Space Telescope\/} Wide Field and Planetary
Camera~2 (WFPC2).  Data taken in the F336W, F439W, and F555W filters
(approximately $U$, $B$, and~$V$) are used to study the surface density
distribution of the $\sim3\times10^4$~stars detected in a 5~arcmin$^2$ region
within $r<2'$ (6.7$\,$pc) of the cluster center.  Realistic simulated images
have been used to estimate photometric errors and incompleteness in the star
counts, which are strong functions of stellar brightness and radius.  We have
used a combination of point-spread-function fitting and aperture photometry,
a technique that yields more accurate photometry than either method alone on
the undersampled WFPC2 images of crowded star fields.  The error in
photometry is $1\sigma\la0.05$~mag for stars with $V<18$; this increases to
$1\sigma\sim0.2$~mag at $V=20.5$, which is 1.5~mag fainter than the main
sequence turnoff.  The surface density of stars in M15 (after correction for
the effects of incompleteness and photometric bias/scatter) is well
represented by a power law in radius: $N(r)\sim{r}^{-0.82\pm0.12}$, over the
radial range 0\farcs3 (0.017$\,$pc) to $6''$.  The observed power law is
remarkably similar to what is expected if the center of the cluster harbors a
massive black hole.  Non-parametric estimates of the density profile show a
monotonic rise with decreasing radius all the way in to $r=0\farcs3$, the
smallest radius at which the density can be reliably measured; there is no
indication that the profile flattens at smaller radii.  Any flat core of
radius larger than $2''$ (0.11$\,$pc) in the stellar distribution is ruled
out at the $\ga95$\% significance level.  The star count profile is
consistent with that expected
from core-collapse models or with the predicted distribution around a massive
(few times $10^3\,M_\odot$) black hole.  The close triplet of bright stars,
AC~214, is within $0\farcs5$ ($1.5\sigma$) of the cluster centroid position.
The projected density distribution of stars within the central $15''$ of M15
departs from circular symmetry at the 95\% level, with an ellipticity
$e=0.05\pm0.04$ (90\% confidence limits) at a position angle of
$+60^\circ\pm25^\circ$, consistent with the rotation measured by Gebhardt
\etal\ (1995) in this region of the cluster.

\vfill\eject

\centerline {\smcap 1. introduction}
\medskip

Galactic globular clusters are excellent laboratories for studying the
dynamics of dense stellar systems.  Theoretical investigations indicate that
close binary stars play a critical role in cluster cores, providing energy to
prevent or halt core collapse and to drive post-core-collapse expansion
(cf.~Pryor \etal\ 1989; Heggie \& Aarseth 1992; Hut \etal\ 1992a, 1992b).
Stellar interactions, including direct collisions and resonant encounters
involving binaries, are thought to be common in dense clusters (Benz \& Hills
1987; Leonard 1989; Leonard \& Fahlman 1991; Hut \etal\ 1992a).  Massive
black holes would concentrate stars in their vicinity and cause the velocity
dispersion to increase (Bahcall \& Wolf 1976, 1977).  It is
difficult, however, to obtain direct observational constraints on the above
processes because of the complications caused by the extreme crowding
near the centers of high concentration star clusters.

Although there have been a number of investigations of the cores of dense
Galactic globular clusters using {\it Hubble Space Telescope\/} ({\it HST\/})
images [Lauer \etal\ 1991; Paresce \etal\ 1991; Ferraro \& Paresce 1993;
Guhathakurta \etal\ 1992 (hereafter referred to as Paper~I), 1994; DeMarchi
\& Paresce 1994; Yanny \etal\ 1994a (hereafter Paper~II)], most published
measurements have been based on data taken with aberrated optics; the size
and complex nature of the Point Spread Function (PSF) made it difficult to
perform accurate photometry in the inner few arcseconds of the cluster (see
above references).  The post-repair {\it HST\/} data are dramatically
improved---the performance of the Wide Field and Planetary Camera~2 (WFPC2)
is described by Trauger \etal\ (1994) and we have used this instrument in a
recent study of the dense cluster M30 (Yanny \etal\ 1994b, hereafter
Paper~IV).  In this paper, we use WFPC2 images of M15 (NGC~7078) to
investigate the stellar density distribution near the center of the cluster.
A study of the stellar populations in M15 will be presented later.
The photometric error with WFPC2 in the crowded central $1'$ of M15 is a
factor of two smaller, and the stellar detection threshold about 2.5~mag
fainter, than in comparable exposures with the pre-refurbishment {\it HST\/}
(Paper~II).

About 20\% of all globular clusters possess what has been termed
`post-core-collapse' morphology, in which the surface brightness appears to
rise all the way into the cluster center; M15 belongs to this category (King
1975; Djorgovski \& King 1984).  Numerical $N$-body simulations (cf.~Heggie
1985) and Fokker--Planck calculations (Grabhorn \etal\ 1992) of core collapse
indicate that the resulting projected density profile slope should be between
$-0.5$ and $-1.5$.  Heating by binary stars can stabilize core collapse
(Heggie 1975; Goodman \& Hut 1989; McMillan \etal\ 1991) and the core is
expected to undergo oscillations even in the absence of binaries (Sugimoto \&
Bettwieser 1983; McMillan 1989; Murphy \etal\ 1990).  These processes are
expected to produce a flattening of the stellar density within a region of
radius a few percent of the half-mass radius, or 0.05--0.1~pc for a typical
globular cluster (Goodman 1989; Gao \etal\ 1991).  This range of core radii
corresponds to $1''$--$2''$ at the distance of M15.  [We adopt $D=11.5$~kpc
as the distance to M15, based on its horizontal branch apparent magnitude
$V_{\rm HB}=16.19$ and a line-of-sight reddening of $E_{B-V}=0.11$ (Fahlman
\etal\ 1985).] The high angular resolution of the post-repair {\it HST\/}
data makes it possible to probe the distribution of stellar surface density
on subarcsecond scales and to measure (or place strong constraints on) the
size of M15's core.

In addition, the high central density of M15, one of the highest of known
globular clusters (Webbink 1985; Djorgovski 1993), has long made it a
candidate for a cluster that has a central black hole (Bahcall \etal\ 1975).
A massive compact object in the center of a cluster would produce a surface
density `cusp' in the steady-state radial distribution of stars, $\sigma(r)$,
with a slope $\alpha\equiv{\rm d(log}\,\sigma)/{\rm d(log}\,r)\sim -0.75$
(Bahcall \& Wolf 1976, 1977), in good agreement with the measured slope (as
we shall see later).  Unless explicitly stated otherwise, we use the term
`cusp' in this paper in a phenomenological sense to describe a density
profile that is increasing roughly as an inverse power law of the distance
from the center of the cluster.  We do not imply by this convenient
representation of the data that any particular theoretical solution is
preferred.

Lauer \etal\ (1991) used an early, pre-refurbishment {\it HST\/} $U$ band
image of the cluster to study the radial distribution of stars.  They found
that the diffuse residual light in M15---the component that remains after
subtraction of the light of bright, resolved stars---appears to flatten near
the cluster center with a core of radius $r_{\rm core}=2\farcs2$.  However,
analysis of realistic simulated images (Paper~II) showed that it is difficult
to interpret the significance of the diffuse light in images of M15 obtained
with the aberrated {\it HST}.  This uncertainty arises because the software
used in the analysis overestimates the brightnesses of stars near the crowded
cluster center (as a result of blending) and because of the broad wings of
the PSF ($r_{\rm PSF}\sim2\farcs5$).  Star counts derived from pre-repair
{\it HST\/} data were consistent with an $\alpha\sim-0.8$ cusp or with an
$r_{\rm core}\la2''$ core.  Our previous limit on $r_{\rm core}$ (Paper~II)
did not take into account the covariance between $r_{\rm core}$ and the
asymptotic power
law index $\alpha$.  Moreover, the systematic effects of a non-uniform
stellar detection threshold in the aberrated data made it difficult to assign
reliable confidence limits to the density profile.  The WFPC2 images of M15
yield a larger and cleaner sample of stars that rule out a core as large as
$r_{\rm core}=2''$ at the 95\% level; the data are well fit by an
$\alpha=-0.8$ power law distribution (see Sec.~5).

Even with the $0\farcs1$ resolution of WFPC2 images, our study of M15's
inner density profile is limited by small number statistics in the star
counts because photometric scatter/bias and incompleteness prevent us from
probing much below the main sequence turnoff.  This fact requires us to use
detailed and realistic simulations to investigate various possible systematic
errors.  While the usual artificial star tests yield an estimate of the
fraction of stars detected, a more extensive simulation (such as that
described in Sec.~4) is needed to simultaneously correct for both
incompleteness and photometric error.

The shape of the stellar density profile near the center of M15 is similar to
the predicted distribution around a massive compact object (Bahcall \& Wolf
1976).  This does not, however, prove that a massive object is present, since
the observed profile slope is also within the range of slopes expected for
collapsed cores.  The definitive way to distinguish between black hole and
core-collapse models is to measure the dispersion in stellar radial
velocities, $\sigma_v$, as a function of radius.  In the presence of a black
hole, $\sigma_v$ is expected to increase inwards roughly as $r^{-0.5}$
(Keplerian dependence).  If, on the other hand, M15 is undergoing core
collapse and does not contain a massive central object, $\sigma_v$ is
expected to be roughly constant over the inner $5''$ at the isothermal value
of $\la15$~km~s$^{-1}$ (Heggie 1985).  The ground-based velocity data of
Peterson \etal\ (1989) are suggestive of a steep rise in $\sigma_v$ within
a few arcminutes of M15's center, exceeding $20$~km~s$^{-1}$ in the inner
$6''$, consistent with that expected for a central black hole of mass
$10^3\,M_\odot$ (Bahcall \& Ostriker 1975; Bahcall \& Wolf 1976, 1977).
However, the inward rise in $\sigma_v$ in the Peterson \etal\ data may not be
significant because of the large sampling errors associated with such a
measurement (Dubath \etal\ 1994; Dubath \& Meylan 1994).  With corrective
optics (COSTAR) installed on {\it HST\/}, it is now possible (though time
consuming) to obtain velocities, accurate to a few km~s$^{-1}$, of tens of
individual post-main-sequence stars in the central $1''$ of M15.  Such data
should settle the debate over whether core collapse or a massive black hole
causes the stellar concentration at the center of M15.

The observations and relevant characteristics of the data are described in
Sec.~2.  A new analysis technique, developed specifically for WFPC2 images,
is outlined in Sec.~3.  In Sec.~4, we describe a simulation that is used to
estimate photometric accuracy and degree of completeness in the data.  [A
reader who is not interested in the details of the procedures used to derive
stellar photometry, error estimates, and star count correction factors may
wish to skip Sec.~3 and 4.]
In Sec.~5, we determine the stellar surface density in M15 as a function of
distance from the cluster center using parametric and non-parametric
techniques, and test for circular symmetry in the star count distribution.
In Sec.~6, we discuss the implications of the results.  A summary is
presented in Sec.~7.

\bigskip
\centerline {\smcap 2. observations}
\medskip

A set of 8~WFPC2 images of the center of M15 was obtained on
a single orbit on 1994 April~7: $4\times8\,$s with the F555W filter,
$2\times40\,$s in F439W, and $2\times200\,$s in F336W.  The pointing of the
telescope was identical for all 8~exposures.  The zero points of the WFPC2
instrumental magnitudes in F336W, F555W, and F785LP have been adjusted so
there is rough agreement ($\la0.1$~mag) with ground-based $UBV$ measurements
of stars in M15 (Stetson 1994); they are also consistent with the in-orbit
WFPC2 calibration (Burrows 1994).  We have opted in favor of empirical
photometric calibration since the instrumental calibration coefficients
change with time, particularly after decontamination events, and are somewhat
dependent on the exact set of flat fields used in the processing pipeline.
In the rest of this paper, we use the terms $U$, $B$, and $V$ interchangeably
with the WFPC2 filter designations when referring to the calibrated
instrumental magnitudes.  The main sequence turnoff, defined to be the bluest
point in the main sequence, is at $V_{\rm TO}=19.1$.  Since our analysis of
the density distribution of M15 relies only on measurements of stellar
brightness {\it relative\/} to $V_{\rm TO}$, and since stellar mass changes
very gradually with $V$~magnitude across $V_{\rm TO}$ (Bergbusch \&
VandenBerg 1992), uncertainties in the photometric zero points or in the
definition of the turnoff point are unimportant for this study.

Each WFPC2 image consists of a mosaic of four $800\times800$ Charge Coupled
Device (CCD) frames---a Planetary Camera CCD (PC1) with a scale of
0\farcs046~pixel$^{-1}$, and three Wide Field CCDs (WF2--WF4) with scales of
0\farcs10~pixel$^{-1}$.  The usable field of view is $34''\times34''$ for PC1
and about $76''\times76''$ for each of the WF CCDs; the total image area is
approximately 5.1~arcmin$^2$.  A greyscale (negative) representation of the
full four-CCD mosaic of the
combined $V$ band image ($4\times8\,$s) of M15 is shown in Figure~1.  The
telescope pointing was chosen so that the cluster center was imaged near the
center of PC1 (upper right quadrant).  The WF2--WF4 images are oriented
counterclockwise starting from upper left.  The gaps between the usable
fields of view of adjacent CCDs ($\la1''$ in width) are not shown to scale in
Figure~1 and the small inter-CCD rotations ($\la1^\circ$) are ignored in this
representation.  A detailed description of the instrumental parameters and
in-orbit characteristics of WFPC2 is given by Burrows (1994) and by Trauger
\etal\ (1994).

The exposure times are short enought to ensure that the brightest stars in
the image ($V\approx12.5$, $B\approx13.5$, $U\approx12.5$) are not saturated
by more than a factor of~4--5 in the individual exposures.  At most 10~pixels
are affected by saturation at the centers of each of these bright stars (more
typically 3--4~pixels), and about 10~stars are affected on each CCD frame,
but it is possible to recover their magnitudes using information in the PSF
wings (Sec.~3).

Figure~2 is a ``true color'' reproduction of the central $9''\times9''$
portion of M15 imaged on PC1.  The blue, green, and red intensities are
roughly proportional to the logarithm of the flux in the $U$, $B$, and $V$
bands, respectively.  The cluster center is marked by the green `$+$' sign
(see Sec.~{\it{5.1}\/}).  This section of the image contains
623~post-main-sequence stars with $V<19$ and over 1600~stars with $V\la21$
which are detected in all three bands, but star identifications are
incomplete for $V>19$ near the center of the cluster.  The average surface
density of post-main-sequence stars ($V<19$) in the central
$9''\times9''$ region shown in Figure~2 is 7.4~arcsec$^{-2}$, which
corresponds to 2370~pc$^{-2}$; the mean surface density within $r<1''$ is
2.4~times higher than this.  The high angular resolution of {\it HST\/}
images is crucial for the study of such crowded regions, especially for
obtaining reliable photometry of stars fainter than the main sequence
turnoff.

\bigskip
\centerline {\smcap 3. analysis technique}
\medskip

We have developed and tested procedures for deriving stellar photometry from
{\it HST\/} images of dense globular clusters.  The reader is referred to
Papers~I and II for a detailed description.  Even though the procedures were
originally designed for pre-repair {\it HST\/} data with an aberrated PSF,
several of the steps are relevant for analysis of crowded star
fields and we apply them here.

Standard bias and flat field calibrations have been applied to the data as
part of the preprocessing pipeline at the Space Telescope Science Institute.
The availability of multiple images in each band, with identical pointings
(pointing offsets $<0\farcs01$) and exposure times, facilitates removal of
cosmic ray events and combination of images, without the need for fractional
pixel interpolation.  While this observing strategy is essential for the
elimination of cosmic rays, it results in a higher effective read noise in
the combined image than if the data were obtained as a single long exposure.
The detectability and photometry of faint stars, being read noise limited in
sparse regions of the image, are thus slightly degraded.  The four short $V$
band exposures from the first set have been median filtered to remove cosmic
rays.  In each of the $B$ and $U$ bands, the difference between the pair of
exposures has been used to identify and mask positive excursions in the
individual exposures (see Paper~IV for details).  The individual cosmic
ray-cleaned exposures in each band have been averaged excluding masked
pixels.  All subsequent data analysis makes use of these cosmic ray-free,
combined $U$, $B$, and $V$ images.

The stellar positions in the $U$ and $B$ band PC1 images are offset from
those in the $V$ band PC1 image by 1.1~pixel and 0.5~pixel
(0\farcs05 and 0\farcs02), respectively.  For the WF2--WF4 CCDs, the
offset between the $U$ and $V$ image ranges from 0.3~pixel to 0.5~pixel
(0\farcs03--0\farcs05) and it is about 0.2~pixel (0\farcs02) between
$B$ and $V$.  The combined images in the three bands have been aligned by
fractional pixel interpolation and summed to produce a deep image.  The
undersampling of the PSF, particularly in the WF data, makes interpolation
somewhat unreliable; however, the sum of interpolated images is only used for
finding stars and not for stellar photometry.  For a typical $V\sim21$
main sequence star, the number of counts (i.e., detected photons) in the $U$,
$B$, and $V$ images is about the same: the system throughput and intrinsic
stellar flux are lower in $U$ (and to a lesser extent in $B$) than in $V$,
but this is
roughly compensated by the longer exposure times at shorter wavelengths.
Coadding the $UBV$ images lowers the detection threshold for faint main
sequence stars by almost 0.6~mag compared to images in a single band.

The standard matched filter peak finding algorithm {\smcap find} of the
{\smcap daophot} software package (Stetson 1987) is used to detect stars on
the summed $U+B+V$ image.  A preliminary round of PSF fitting is done using
this star list, a library PSF template, and the {\smcap allstar} program in
{\smcap daophot} and the residual image (original$\,-\,$template) is
inspected.  The star list is then manually edited to separate blended
stars (typically a few percent of the stars in the list), to add faint stars
whose peak brightnesses lie below the {\smcap find} detection threshold, and
to remove hot pixels and PSF artifacts mistakenly identified as stars.  The
master list contains $1.2\times10^4$~stars from the PC1 image
(0.33~arcmin$^2$) and about $9\times10^3$, $6\times10^3$, and
$1.0\times10^4$~stars from WF2--WF4, respectively (each WF CCD covers an area
of 1.56~arcmin$^2$).

The analysis technique used in this paper incorporates a set of standard
{\smcap daophot ii} routines (Stetson 1992).  The data analysis (PSF
reconstruction, PSF fitting, etc) is done independently for the $U$, $B$,
and $V$ bands, and independently for each of the four CCD frames in the WFPC2
mosaic.  Only the master star list is used as a common starting point for
PSF fitting in the three bands.

The data analysis procedure begins by correcting for saturation in the
central few pixels of the images of the brightest 5--10~stars in each CCD
frame.  A small PSF template ($r_{\rm PSF}=10$~pixels) is constructed from
unsaturated stars, fitted to the unsaturated PSF wings of the saturated stars
(${\rm2~pixels}\la{r}<{\rm4~pixels}$), and the results of the fit are used to
replace all pixels values above the saturation threshold
($\sim2.5\times10^4$~electrons).  A series of iterations is performed on the
saturation-corrected images.  Each iteration step consists of constructing
an empirical PSF template of radius 20~pixels by averaging the images of
about 30~relatively isolated, bright stars, after removing the neighbors of
these stars using the PSF template from the previous iteration.
The quality of the PSF template improves with each iteration as the neighbors
of the PSF stars are removed with increasing precision, reaching convergence
after 3--4~iterations.  Of all
the templates tested---bivariate Gaussian, Lorentzian, and Moffat functions
with a matrix of fixed or linearly changing residuals---the combination of a
Moffat function and residuals which vary linearly with image position
produces the best approximation to the actual PSF.  The final PSF template
is simultaneously fitted to all stars detected on the image using the
positions in the master list as input to the {\smcap allstar} routine in
{\smcap daophot}.

For each star, the results of the {\smcap allstar} PSF template fit are used
to remove every neighboring star within a 40~pixel radius ($2\,r_{\rm PSF}$).
Aperture photometry of each star is then obtained using a circular aperture
of radius 2~pixels for stars on PC1 ($r=0\farcs09$) and 1.6~pixels for stars
on WF2--WF4 ($r=0\farcs16$), with the local sky background measured in a
surrounding annulus.  These aperture radii represent optimal values as
determined from tests conducted on simulated images and on the M15 data, the
latter based on the tightness of the red giant branch (RGB) and main sequence
in a color--magnitude diagram (CMD).  Measurements with smaller apertures
lead to noisier photometry due to the effects of undersampling and
intra-pixel quantum efficiency variations (Burrows 1994), while photometry
from larger apertures are affected to a greater extent by contamination from
poorly subtracted neighboring stars.  We have experimented with three
different photometric techniques on a simulation of the M15 PC1 image
(Sec.~4): direct aperture photometry, PSF fitting, and the hybrid of these
two methods described above.  For bright stars ($V\sim15$), the hybrid method
and aperture photometry yield comparably good results
($1\sigma\sim0.015$~mag) since the effect of neighbors is negligible
(particularly on color measurements), but PSF fitting is significantly less
accurate ($1\sigma\sim0.03$~mag).  In fact, PSF fitting is even less accurate
for bright stars in the WFPC2 image of M15 ($1\sigma\sim0.04$~mag, judging
from the width of the bright RGB) than for those in the simulated image,
because the effect of intra-pixel quantum efficiency variation has not been
included in the simulated image and because the PSF is slightly broader, and
therefore better sampled in the simulation than in the PC1 image (the
broadening is a result of the fact that the simulation PSF template was
derived by interpolating between stellar images on the WFPC2 image of M15).
Direct aperture photometry of faint stars ($V\sim19$--20) is severely
affected by neighbor contamination ($1\sigma\sim0.45$~mag and the mean
photometric bias is 0.25~mag), but the hybrid method and PSF fitting perform
better ($1\sigma\sim0.30$~mag; mean $\rm bias=0.07$~mag).

Photometry of a set of bright, isolated stars in a series of concentric
apertures indicates that the $r=2$~pixel aperture contains about 75\% of the
light contained in an $r=20$~pixel aperture around the $V$ band PC1 PSF.
The enclosed fraction is slightly higher for the $r=1.6$~pixel WF aperture
and for the $U$ and $B$ bands.  Differences in the aperture correction are
accounted for before combining the PC data with data from the WF CCDs.  For a
given band and CCD, the fraction of light within the aperture appears to be
essentially independent of position on the CCD.  The PSF in the M15 PC1 image
is somewhat more centrally concentrated than the one in the PC1 image of M30
which was obtained a week earlier than the M15 data (Paper~IV).  This is
probably caused by orbital variations in the {\it HST\/} focus, commonly
referred to as ``breathing'' (Trauger \etal\ 1994).

\bigskip
\centerline {\smcap 4. simulation}
\smallskip

\centerline {\it 4.1 Construction of the Simulated Image}
\smallskip

The high degree of crowding near the center of M15 greatly complicates the
stellar photometry and star count analysis.  To estimate the errors and
biases introduced by crowding, we have constructed a simulated image and have
analyzed it in the same manner as the M15 data (cf.~Papers~I and II for a
description of this procedure).  The simulated image is designed to mimic the
combined $V$ band M15 PC1 image as closely as possible.  We have not
simulated the WF data; the stellar density (and degree of crowding) is
low enough throughout the WF images to ensure that counts of stars are
essentially complete for $V<20$.

The basic ingredients needed to build the simulated image are:
(1)~the shape of the stellar luminosity function (LF); (2)~the surface
density of stars, brighter than some specified limiting magnitude, as a
function of projected distance from the cluster center; and (3)~the position
dependent shape of the stellar PSF.  As described below, each of these input
parameters is derived from the PC1 $V$ image of M15.

For the purposes of the simulation, the stellar LF is assumed to have the
same shape at all radii.  The {\it HST\/} M15 data show no evidence for
luminosity segregation vs.\ radius.  Figure~3 shows the $V$ band LF as a
function of radius in M15.  The stars detected in the full WFPC2 mosaic are
divided into 8~annuli:
(1)~$r<5\farcs4$,
(2)~$5\farcs4\leq{r}<9\farcs8$,
(3)~$9\farcs8\leq{r}<15''$,
(4)~$15''\leq{r}<24''$,
(5)~$24''\leq{r}<35''$,
(6)~$35''\leq{r}<48''$,
(7)~$48''\leq{r}<66''$, and
(8)~$66''\leq{r}<127''$.
The boundaries of these annuli have been chosen so that they contain roughly
the same number of $V<19$ stars; our sample should be nearly complete at all
radii for stars brighter than this limiting magnitude (see below).  The LF of
post-main-sequence stars in M15 has the same shape at all radii, except for
the slight deficiency of bright stars in the innermost radial bin.  The bump
at $V\sim16$ is caused by horizontal branch (HB) stars.  Artificial star
tests and previous simulations (Papers~II and IV) indicate that the turnover
at the
faint end of the LF is caused by incompleteness, which is a strong function
of distance from the center---the observed LF peaks at $V\sim19$ (the main
sequence turnoff point) in the innermost radial bin, but rises beyond that
and peaks about 2~mag fainter in the outermost radial bin.  The open circles
in Figure~3 show the LF used to construct the simulated image (arbitrary
normalization).  It is derived from the M15 PC1 data, by averaging the data
for bright stars ($V\la19$), and by adopting the upper envelope of the LFs in
the range $V=19$--20.  For $V>20$, the LF is assumed to be a power law with a
slope ${\rm d(log}\,N)/{\rm d}V=0.125$.  The details of the shape of the
synthetic LF beyond $V=20$ are unimportant for the purpose of learning about
faint stars; the photometric accuracy and detection efficiency for $V>20$
stars are determined by the local surface density of brighter stars
($V\la18$) in crowded regions and are limited by read noise and photon noise
in sparse portions of the image.  Furthermore, the faint end slope of the LF
is such that the main sequence luminosity is dominated by stars at the bright
end ($V\sim19$--20) and the contribution from stars fainter than $V=20$ is
negligible.

The projected density of stars in the simulated image is described by a
radial distribution function consisting of two power laws, one with an index
$\alpha_1=-0.7$ within $r<6''$ of the cluster center and the other with an
index $\alpha_2=-1.3$ for $r>6''$.  This is a good approximation to the shape
of the (uncorrected) density profile of bright ($V\la18$) stars in the inner
$30''$ of M15 (see Sec.~5).  The PC1 data should be complete for $V<18$
stars, so their radial distribution is unlikely to be strongly affected by
differential incompleteness as a function of distance from the cluster
center.  The overall normalization of the stellar density in the simulation
is the same as in M15: about 6~arcsec$^{-2}$ at $r=1''$ to a limiting
magnitude of $V=18$.

Each star is assigned a $V$ magnitude based on the synthetic LF, a projected
radius based on the dual power law radial distribution function, and a
random azimuthal angle.  Artificial stellar images are then added to a blank
$800\times800$ frame using a linearly variable PSF template (as defined by
Stetson 1987) derived from the M15 PC1 data.
The center of the simulated cluster is assumed to be near the center of the
frame.  Poisson noise and read noise are added to the simulated image based
on the instrument characteristics of the PC1 CCD and on the length of the M15
$V$ exposures.  The final simulated image is similar (in a statistical sense)
to the combined $V$ band image ($4\times8\,$s) of M15.  This image is
analyzed using the procedure described in the last section.  In the course of
the analysis, we deliberately avoid using any input data that were used in
the construction of the simulated image.  Instead, we use the {\smcap find}
routine to detect stars, reconstruct a PSF template from the detected stars,
use {\smcap allstar} to perform PSF-fitting, and derive an aperture magnitude
for each detected star after subtraction of its neighbors.

\medskip
\centerline {\it 4.2 Photometric Accuracy}
\smallskip

The analysis program derives a list of stellar positions and brightnesses
from the simulated image; this is referred to as the `output list'.  Each
star in the output list is matched to the star closest to it in the `input
list' (the list of positions and magnitudes used to construct the simulated
image), provided their positions agree to within 1~pixel ($0\farcs05$ for
PC1; $0\farcs1$ for the WF images).  Essentially all the stars in the output
list (over $1.1\times10^4$ in number) are successfully matched.  The
difference between the input and output $V$ magnitudes of all matched stars
is plotted against the `true' $V$ magnitude ($V_{\rm input}$) in the top
panel of Figure~4.  The stars tend to be concentrated near the $V_{\rm
output}=V_{\rm input}$ line, but their distribution about it is asymmetric.
The photometric bias, in the sense of the output magnitudes being
systematically brighter than the input ones, is caused by blending in crowded
parts of the image: two or more input stars within $\sim0\farcs05$ of one
another are fit as a single, brighter star.  Some of the extreme values of
$V_{\rm input}-V_{\rm output}$ are mismatches---a small error in astrometry
occasionally leads to a configuration where the input star closest to the
output star is {\it not\/} its true counterpart.  The reader is referred to
Papers~I and II for details.

We divide the sample of matched stars into two parts: stars located
within $5''$ of the center of the simulated cluster (13\% of all matched
stars) and those beyond $5''$.  These two subsamples are presented in
the middle and bottom panels of Figure~4, respectively.  The open circles
show the mean magnitude offset in 0.5~mag bins, and the error bars show the
$1\sigma$ rms dispersion within each bin.  The filled circles indicate the
median value of $V_{\rm input}-V_{\rm output}$, and the error bars include
$\pm34\%$ of the stars on either side of the median (this corresponds to the
fraction of points enclosed by $1\sigma$ error bars around the peak of a
Gaussian distribution).  The dashed lines indicate $\pm10$\% photometric
errors.  The large error in the $V=13.75$ bin in the inner $5''$ (middle
panel) is a result of the fact that one of the two stars in this bin is
blended with a $V=14$ star causing its output magnitude to be in error by
$+0.6$~mag (upper panel).  The asymmetry in the distribution of magnitude
differences (photometric bias) is reflected by the fact that the $+34\%$
errors are larger than the $-34\%$ errors.  It is evident in the top panel of
Figure~4 that the distribution peaks close to zero, and the median values are
less strongly affected by the positive tail of the distribution than the mean
values.

The accuracy with which the analysis program recovers the magnitudes of
individual stars in the simulation depends both on the stellar brightness and
on the local surface density of stars, the latter being a montonically
decreasing function of distance from the cluster center.  In the relatively
sparse outer region, $5''<r<25''$, the $1\sigma$ scatter in photometry is
$\la0.03$~mag for stars with $V<17$, about 0.1~mag at $V=18$, and 0.25~mag at
$V=20$, while the mean bias is $\leq0.1$~mag for $V<20$.  By comparison, the
$1\sigma$ scatter (and mean bias) for stars in the crowded inner $5''$ is
about 0.2~mag ($\ga0.1$~mag) at $V=18$ and 0.4~mag (0.4~mag) at $V=20$.  The
photometric scatter and bias within $r<5''$ increase rapidly beyond $V=20$.
However, only 30\% of the input stars with $V=20$, and less than 10\% of
$V=21$ stars, are detected in this region (see Sec.~{\it{4.3\/}}).
The detected fraction of $V_{\rm input}>20$ stars represents the positive
tail of the $V_{\rm input}-V_{\rm output}$ distribution---those stars for
which blending, errors in sky determination, and photon/read noise have
caused a large enough brightening (by $\ga0.5$~mag) for them to be above the
detection threshold in the crowded inner region.

Our simulation shows that there are two principal sources of photometric
error in the crowded portions of the PC1 image of M15: (1)~blending of two or
more stars into a single apparent object, and (2)~inaccurate subtraction of a
close, but resolved, star.  The first of these leads to a systematic bias in
photometry while the second can cause positive or negative measurement
errors.  For $V>19$ stars within the inner $5''$, the errors caused by
blending appear to be comparable to or larger than those caused by improper
neighbor removal.  The blending biases brightness measurements but not
necessarily color measurements, since it mostly affects main sequence stars
just below the turnoff, all of which have roughly the same color
($B-V\sim0.4$--0.5 for stars in the range $V=19$--22).

\medskip
\centerline {\it 4.3 Correction of Star Counts}
\smallskip

To correct the observed star counts in M15 for incompleteness and photometric
error, the input and output star lists from the simulation are compared.  We
define a star count correction factor:
$$ C ~=~ {{N(V_{\rm output}<V_{\rm lim})} \over {N(V_{\rm input}<V_{\rm
lim})}} ~~~~~, \eqno(1)$$
where the input and output stars are counted over the same area.  It is
important to note that the counts in the numerator and denominator of the
above equation are drawn independently from the output and input lists,
respectively---they are {\it not\/} based on matched star counts.  The
quantity $C$ measures the full `throughput' of the analysis procedure,
including the effects of incompleteness and photometric error, not merely the
detection efficiency.

Figure~5 shows $C$ as a function of radial distance $r$ from the cluster
center, with the stars binned in $0\farcs4$-wide annuli.  Three different
limiting magnitudes, $V_{\rm lim}=18.3$, 19.0, and 20.0, are used in the top,
middle, and bottom panels of Figure~5, respectively.  The error bars
represent the Poisson error in the number of output stars.  While the
non-detection of stars decreases the value of $C$, photometric bias ($V_{\rm
output}<V_{\rm input}$) tends to increase $C$, resulting in values greater
than unity in some cases.  Even in the absence of bias, photometric errors
cause more stars to be scattered into a magnitude-limited sample than out of
it due to the fact that the stellar LF rises towards fainter magnitudes
(Fig.~3).  Non-detection of stars dominates over the effect of photometric
bias/scatter in the inner few arcseconds for $V_{\rm lim}\ga19$.  At large
radii ($r\ga10''$) and for $V_{\rm lim}\la20$, $C$ approaches unity as 100\%
of the input stars are detected and the effect of photometric error on
the star counts become negligible.  In choosing the optimal value of $V_{\rm
lim}$, there is a trade-off between the shot noise associated with the
relatively small number of stars at bright limiting magnitudes (which affects
both the estimation of $C$ and the M15 star counts), and the strong radial
dependence and possible systematic uncertainties in the correction factor at
faint limiting magnitudes.

We use an analytic approximation of the form:
$$C(r) ~=~ 1 ~+~ C_1 e^{-r/r_1} ~+~ C_2 e^{-r/r_2} \eqno(2)$$
to represent the correction factor derived from the simulation (smooth curves
in Fig.~5).  The scale length $r_1$ is set to $5''$ for all values of $V_{\rm
lim}$ and the last term is excluded ($C_2=0$) for $V_{\rm lim}=18.3$.  The
remaining parameters are: $C_1=(0.5,\, 0.8,\, 0.1)$ for $V_{\rm lim}=(18.3,\,
19,\, 20)$, and $C_2=(-0.9,\, -0.85)$ and $r_2=(1\farcs5,\, 2\farcs5)$ for
$V_{\rm lim}=(19,\, 20)$.  The correction factor for the $18.3<V<20.0$ sample
is well fit by the parameters $C_2=-0.95$, $C_1=0$, and $r_2=2\farcs5$.  In
deriving the radial density profile of M15 in Sec.~5, we correct the
magnitude-limited star counts by dividing them by the above formulae for
$C(r)$.

\bigskip
\centerline {\smcap 5. density distribution}
\smallskip

\centerline {\it 5.1 Cluster Center}
\smallskip

The location of the center of M15 is determined by calculating the centroid
of the positions of all stars brighter than $V_{\rm lim}$ within a circle of
radius $r_{\rm lim}$.  The centroid computation is iterative: the center of
the circle used to define the sample is set to the centroid calculated in the
previous iteration.  The process converges within 10 iterations for initial
guesses within a few arcseconds of the obvious cluster center.  Each star is
given equal weight in the centroid calculation; unlike estimates based on the
surface brightness distribution, a number-weighted centroid is not biased
towards the few bright RGB stars that dominate the cluster light (cf.~the
shape of the bright end of the LF in Fig.~3).  The centroid is insensitive to
the choice of limiting radius and magnitude for $5''<r_{\rm lim}<15''$ and
for $20<V_{\rm lim}<21$.  Smaller values of $r_{\rm lim}$ and $V_{\rm lim}$
yield samples containing too few stars to permit accurate determination of
the centroid.  The mean of the centroids of 10 stellar samples (various
combinations of $r_{\rm lim}$ and $V_{\rm lim}$) is adopted as the cluster
center.  Its coordinates, relative to the bright reference star AC~211, are:
$$\Delta\alpha_{2000}\,({\rm center})= +0\farcs36\pm0\farcs2 \, ; ~~~
 \Delta\delta_{2000}\,({\rm center})= -1\farcs97\pm0\farcs2 \eqno(3{\rm a})$$
(see Paper~II for an explanation of this coordinate system), and the
coordinates of AC~211 are given by Kulkarni \etal\ (1990):
$$~~~~~~\alpha_{2000}\,({\rm AC~211})= 21^{\rm h}29^{\rm m}58\fss26 \, ; ~~~
 \delta_{2000}\,({\rm AC~211})= +12^\circ10^\prime02\farcs9 ~~~~~~.
 \eqno(3{\rm b})$$

The rms scatter among the centroids of the 10 samples is
$\delta{r}\sim0\farcs3$ (6~pixels on the PC1 image).  The error in the
centroid of each sample is estimated using the bootstrap resampling method
(Efron 1982); it is typically $0\farcs2$--$0\farcs3$.  Thus, a conservative
measure of the $1\sigma$ error in the determination of the cluster center is
$0\farcs3$.  If the center we have adopted is off the ``true'' cluster center
by an amount $\Delta{r}$, it will cause a spurious flattening in our estimate
of the density profile interior to $\Delta{r}$.  We restrict our analysis of
the stellar density distribution in M15 to radii larger than the $1\sigma$
positional error in the cluster centroid.

There appears to be a local maximum in the projected density of stars
about $0\farcs3$ to the north-west of the cluster center adopted here, within
the $1\sigma$ error circle for the center.  This can be seen as a
concentration of faint stars in Figure~2 immediately above the green `$+$'
marking the mean centroid position.  Merritt \& Tremblay (1994) derived
adaptive kernel estimates of the stellar surface density in M15 using
pre-refurbishment {\it HST\/} data from Paper~II, and found a peak at exactly
this location, offset to the north-west of the centroid position (see Fig.~12
of their paper).  We have {\it not\/} adopted the position of the density
peak as the cluster center (except for using it as an alternate center) as
this would bias our estimate of the radial density profile in the sense of
making it as steep as possible near $r=0$.
It is interesting that the object AC~214 (Auri\`ere \& Cordoni 1981), which
was resolved into a close triplet of horizontal branch stars using early {\it
HST\/} images (Paper~II), is located about $0\farcs5$ ($1.5\sigma$) east of
the centroid position.

\medskip
\centerline {\it 5.2 Star Counts vs.\ Radius}
\smallskip

Determination of the radial density profile near the center of M15 is the
principal motivation for obtaining these high resolution images.  Luminous
stars serve as a convenient tool for measuring the surface mass density.
Ground-based data, such as the $UBV\!R$ images of M15 obtained by Lugger
\etal\ (1987) in $1\farcs5$ seeing, or even Stetson's (1994)
$0\farcs5$-seeing images, lack the resolution necessary to study sufficient
numbers of individual stars near the dense cluster center.  The disadvantage
of using the overall surface brightness distribution (Lugger \etal\ 1987) is
that it is dominated by a few stars at the tip of the RGB which are not
representative of the cluster's stellar population, not even its
post-main-sequence population---{\it e.g.}~most of the light within $r<1''$
of M15's center comes from only 7~stars.  It is customary to subtract the
contribution of the brightest few stars and to analyze the residual diffuse
light (cf.~Lauer \etal\ 1991).  In clusters as dense as M15 though, the
residual light is greatly affected by inaccurate photometry and removal of
the bright stars, the errors being largest near the crowded cluster center
(see Sec.~{\it{3.4\/}} of Paper~II for a detailed discussion).

Counts of individual stars are a fairer tracer of the stellar mass density
than the luminosity-weighted star distribution, since all post-main-sequence
and turnoff stars have roughly the same mass ($M\sim0.8\,\msol$) despite the
fact that their visual luminosities span over two orders of magnitude.  Mass
segregation is unimportant for stars brighter than $V\sim20$---the Bergbusch
\& VandenBerg (1992) isochrone for M15 predicts a stellar mass of
$M=0.77$--0.79$\,M_\odot$ for $V\leq19$ (above the main sequence turnoff),
falling only slightly to $M=0.74\,M_\odot$ at $V=20$.  Magnitude-limited star
counts are, however, affected by incompleteness and photometric errors; we
correct for these effects with simulations of the data.

We have compared star counts derived from WFPC2 data in the inner region of
M15 to those derived from pre-repair HST Planetary Camera data (Paper~II).
There are 58~stars in the WFPC2 list with $V<19.0$ (and a total of 82~stars
down to $V=19.6$) within $1''$ of the cluster center, compared to 28~stars
with $V<19.0$ (and a total of 33~stars down to $V=19.5$) in the pre-repair
list for the same region.  The two datasets contain 489~stars and 232~stars,
respectively, between $1''<r<3''$, and 1100~stars and 523~stars,
respectively, between $3''<r<6''$, down to a limiting magnitude of $V=20$.
The WFPC2 sample of stars in the crowded central region of M15 is more than
double the size of the pre-repair sample, and has a more uniform degree of
completeness than the latter.

Figure~6 shows the radial distribution of star counts for two independent
magnitude-limited samples, $V<18.3$ and $18.3<V<20.0$, within $5''$ of the
cluster center in $0\farcs4$-wide radial bins.  The dashed lines show raw
(uncorrected) counts while the open squares show corrected counts, using the
analytic approximation to the correction factor $C(r)$ given by Eq.~(2).  The
shapes of the corrected star count profiles for the two samples are
consistent with one another to within Poisson errors (error bars in Fig.~6),
even though the uncorrected profiles and the $C(r)$ curves derived from the
simulated image are quite different for the two.  The radial profile shapes
of the samples are {\it expected\/} to be the same---the average mass of the
stars comprising the $V=18.3$--20 sample is $M\sim0.75\,M_\odot$ compared to
$M=0.78\,M_\odot$ for the $V<18.3$ sample; this difference in mean stellar
mass is too small to expect significant effects due to mass segregation
(cf.~Bolte 1989; Pryor \etal\ 1986).  The similarity in the shapes of the
corrected density profiles suggests that the correction factors derived from
the simulation are accurate.  The counts in the $V_{\rm lim}=18.3$ sample are
actually corrected {\it downward}; this is due to the effect of photometric
errors on magnitude-limited counts discussed in Sec.~{\it{4.3}}.  The
$18.3<V<20.0$ counts, on the other hand, are increased because incompleteness
effects dominate over the error introduced by photometric bias.

For the purpose of making density measurements, we group the stars in M15
into three samples with limiting magnitudes of $V_{\rm lim}=18.3$, 19.0, and
20.0.  Magnitude-limited samples are usually dominated by stars with
$V\approx{V_{\rm lim}}$ since the typical stellar LF rises steeply towards
the faint end (Fig.~3).  For example, at $r=5''$, the $V<18.3$ sample
constitutes less than 45\% of the $V<19.0$ sample and only 16\% of the
$V<20.0$ sample.  Interior to $0\farcs4$, however, the high degree of
crowding causes most stars fainter than $V=18.3$ to go undetected, implying
that the three samples are nearly identical in this region.  The star count
correction factor $C$ within $r<0\farcs4$, on the other hand, is quite
different for these samples (see Fig.~5).  Even though the raw counts in the
three samples are {\it not\/} independent of one another, the corrected star
count profiles derived from them provide complementary measures of the
density of M15.  The use of a bright limiting magnitude ($V_{\rm lim}=18.3$)
has the advantage that $C(r)$ is not a strong function of radius---it changes
by less than 30\% within $r<5''$.  The disadvantage is that Poisson errors in
the star counts are relatively large.  The reverse is true for the $V_{\rm
lim}=20.0$ sample: it is limited not by Poisson statistics but by
uncertainties in the count correction factor coupled with the fact that $C$
departs strongly from unity near the cluster center.  The $V_{\rm lim}=19.0$
sample represents a reasonable compromise between these two sources of
uncertainty.

Figure~7 shows the corrected $V_{\rm lim}=19.0$ and 20.0 stellar density
profiles (open squares and `$\times$', respectively) with $1\sigma$ Poisson
error bars based on the raw counts.  The $V<20.0$ counts have been shifted
down by~0.6~dex to match the normalization of the $V<19.0$ sample.  The data
in the range $0\farcs3<r<5''$ are consistent with a power law density
profile, $\sigma(r)\sim{r}^\alpha$, with the index $\alpha=-0.82$
(curve~1).  A core of radius $r_{\rm core}=1''$ (0.06~pc) or smaller is
allowed by the data but a $2''$ core appears to be ruled out.  Note that
curves~2 and 3 in Figure~7 are of the form:
$$ \sigma(r) ~~=~~ \sigma_0 \, [ \, 1 ~+~ C_\alpha (r/r_{\rm core})^2 \,
   ]^{\alpha/2} ~~~~~~, \eqno(4{\rm a})$$
where
$$ C_\alpha ~=~ 2^{-2/\alpha} \, - \, 1 ~~~~~~. \eqno(4{\rm b})$$
This form of $\sigma(r)$ approaches a power law with index $\alpha$ for
$r>\!\!>r_{\rm core}$, while the core radius $r_{\rm core}$ has the usual
property that the surface density at this radius is half the central density
$\sigma_0$.  The asymptotic index $\alpha$ is set to $-0.8$ for curves~2 and
3, the value that best fits M15 data in the range $r=2''$--$5''$.

When comparing the M15 star count profile to model curves, it should be noted
that the estimate of the stellar density within $r<0\farcs3$ (to the left of
the dotted vertical line in Fig.~7) is likely to be biased low due to errors
in the determination of the cluster center.  To test how sensitive the
estimate of the density profile is to the choice of cluster center, we have
adopted the location of the stellar concentration $0\farcs3$ to the
north-west of the cluster centroid (Fig.~2; Sec.~{\it{5.1}}) as an alternate
cluster center.  The bold dashed line in Figure~7 shows the corrected density
profile of $V<19.0$ stars in M15 about this alternate center; it is
consistent with the profile measured relative to the centroid (open squares)
to within Poisson errors.

Stars from the entire WFPC2 mosaic image of M15 are used to determine the
stellar surface density distribution from the center out to $r=2'$.  This
radial density profile is plotted in Figure~8.  Only $V<19$ stars are shown,
with no correction for incompleteness/photometric error.  These corrections
are small for $V_{\rm lim}=19$ and for $r\ga5''$.  The uncorrected profile is
well fit by a combination of three power law sections: an inner index
$\alpha_1=-0.7$ within $r<6''$; $\alpha_2=-1.3$ in the range $6''<r<30''$;
and $\alpha_3=-2$ for $r>30''$.  The stellar surface density profile beyond
$r=30''$ in M15 follows the $r^{-2}$ slope (dotted line) expected for
$r>\!\!>r_{\rm core}$ in an isothermal profile.  The solid lines in Figure~8
indicate the radial distribution function used to construct the simulated
image (Sec.~{\it{4.1\/}}).  The apparent dip in the profile at $r=13''$ may
be a result of inaccurate photometry of stars imaged near the gap between PC1
and the WF CCDs.

\medskip
\centerline {\it 5.3 Parametric Fits to the Radial Density Distribution}
\smallskip

In this section, we compare the radial distribution of stars in the inner
$5''$ of M15 to various model profiles.  The model profiles have been
multiplied by the correction function $C(r)$ [Eq.~(2)] in order that they may
be fitted directly to the raw (uncorrected) magnitude-limited counts.  Unlike
direct estimates of the stellar density for which the data must be divided by
$C(r)$ (a procedure that can be uncertain if $C$ is much less than unity),
this approach attaches low weight to the model in regions where $C$ is small.
Another attractive feature of parametric model fitting is that the stars do
not need to be binned into annuli.

Figure~9 shows the cumulative radial distribution $f_{\rm cum}(r)$ of $V<19$
stars in M15 (thin stepped line):
$$ f_{\rm cum}(r) ~=~ {{N(r_{\rm min}<r^\prime<r)} \over
   {N(r_{\rm min}<r^\prime<r_{\rm max})}} ~~~~~~. \eqno(5)$$
The minimum radius $r_{\rm min}$ is set to $0\farcs3$, the $1\sigma$
positional error in the cluster centroid, in order to restrict the sample to
stars for which the radial distance is reasonably well determined.  We adopt
an outer boundary of $r_{\rm max}=5''$ since we are interested in the shape
of the density profile near the center of the cluster.  The smooth bold curve
in the upper panel shows the best fit power law while the bold dashed curves
represent profiles which have only a 1\% Kolmogorov--Smirnov probability of
being consistent with the data: $\alpha=-0.82\pm0.12$.  The bold dashed
curves in the lower panel of Figure~9 are profiles with $r_{\rm core}=1''$
(upper curve) and $r_{\rm core}=2''$ (lower curve), with the asymptotic index
$\alpha$ for both models set to $-0.82$ [Eq.~(4)].  Neither of these models
is as good a fit to the data as the $\alpha=-0.82$ profile with $r_{\rm
core}=0$ (bold solid line in upper panel).  This does not, however, imply
that the presence of a $1''$ core is ruled out.  There is finite covariance
between $r_{\rm core}$ and $\alpha$; this implies that the two parameters
must be fitted simultaneously to the data in order to derive a meaningful
estimate of the allowed range of parameter space.

Using the functional form given in Eq.~(4), we carry out a two parameter
($r_{\rm core}, ~\alpha$) maximum likelihood fit to the M15 star count
distribution.  We fit the $V<19.0$ and $V<20.0$ samples separately,
multiplying the model by the corresponding star count correction factor
$C(r)$ in each case [Eq.~(2)].  Standard $\chi^2$ tables for two parameter
functional fits are used to derive error estimates for $r_{\rm core}$ and
$\alpha$.  The results of the maximum likelihood analysis are shown in
Figure~10 in the form of 50\%, 75\%, 90\%, 95\%, and 99\% confidence
contours.  These confidence levels do not take into account uncertainties in
the determination of the correction factor.  There is a significant degree of
covariance between the core radius $r_{\rm core}$ and the asymptotic power
law index $\alpha$ for both samples.  The $V<20.0$ data are consistent with a
pure power law profile ($r_{\rm core}=0$), with a best fit index of:
$$~~~~~~~~~~~~~~\alpha=-0.82\pm0.12 ~~~~~~\rm (95\%~confidence~limits) ~~.
  \eqno(6{\rm a})$$
A profile with a flat core of radius:
$$~~~~~~~~~~~~r_{\rm core}<2''~\rm (0.11~pc)~~~~~~(95\%~upper~limit)
 \eqno(6{\rm b})$$
can also be accommodated by the data but only by making the asymptotic
index $\alpha$ steeper---i.e., more negative---than $\sim-1$.  The 90\% upper
limit on $r_{\rm core}$ is $1\farcs8$ (0.10~pc).  The $r_{\rm core}=2\farcs2$
profile, which appeared to be a good fit to the diffuse $U$-band residual
light in the deconvolved pre-repair HST image (Lauer \etal\ 1991), is ruled
out that the 99\% level by WFPC2 star counts.

The allowed region of $r_{\rm core}$--$\alpha$ space is roughly similar for
the $V_{\rm lim}=19.0$ and $V_{\rm lim}=20.0$ samples.  The $V_{\rm
lim}=20.0$ sample contains about 2.5~times as many stars (raw counts) as the
$V_{\rm lim}=19.0$ sample between $2''<r<5''$, so that the index $\alpha$ is
better constrained.  The limits on $r_{\rm core}$ derived from the two
samples are comparable: the $V<20.0$ sample contains only 40\% more stars
(raw counts) than the $V<19.0$ sample within $r<2''$, and its $C(r)$ curve
drops sharply with decreasing radius in this region (see Fig.~5), implying
that the innermost $2''$ gets low relative weight in the maximum likelihood
fit.  Errors in estimating $C(r)$ have a bigger effect on the $V<20.0$ sample
than on the $V<19.0$ sample since the correction factor is closer to unity
for the latter sample.

\medskip
\centerline {\it 5.4 Non-parametric Estimates of the Radial Density Profile}
\smallskip

In the last two sections, the shape of the density profile was estimated by
binning the stars in M15 in $0\farcs4$-wide radial bins and by fitting model
profiles to the radial distribution of stars.  Merritt \& Tremblay (1994)
argue in favor of non-parametric methods for estimating the density profile
over the above techniques since the stars need not be grouped (binning
introduces an arbitrary degree of smoothing into the data) and it is not
necessary to compare the data to specific, parameterized model density
profiles.  These authors describe the non-parametric MAximum PEnalized
Likelihood (MAPEL) scheme which merely requires that the profile be smooth
`locally' (in the radial domain).  The degree of smoothness is specified by
the $\lambda$ parameter; the larger the value of $\lambda$, the more closely
does the estimate approach a `smooth solution'.  The penalty function, which
defines what is meant by a `smooth solution', is a power law in radius.
Thus, in the limit $\lambda\rightarrow\infty$, one obtains a single power
law, maximum likelihood fit to the star counts.  Ideally, $\lambda$ should be
large enough so that the solution is smooth, but not so large as to bias the
solution.  The number of data points in the sample may be used to determine
the optimal value of $\lambda$ (Thompson \& Tapia 1990; Merritt \& Tremblay
1994).

We have applied the MAPEL method to the stellar distribution in the inner
$5''$ of M15.  In Figure~11, we present non-parametric estimates of the
surface density profile derived from the $V_{\rm lim}=18.3$, 19.0, and 20.0
samples.  The dashed lines are based on uncorrected counts while the solid
lines are based on applying the correction factor $C(r)$ measured from the
simulation (Sec.~{\it{4.3\/}}).  Stars within $0\farcs3$ of the cluster
centroid, for which the radial distance and correction factor are uncertain,
are excluded from the analysis.  Note the striking similarity in the shapes
of the corrected profiles, despite the differences between the uncorrected
ones.  Smoothing parameters of $\lambda=1\times10^{-4}$, $5\times10^{-5}$,
and $2\times10^{-5}$ are close to optimal for the $V<18.3$, $V<19.0$, and
$V<20.0$ samples, respectively (Merritt, private communication).

The non-parametric estimates of the surface density profile in M15 can be
used to check the validity of the parametric fits.  In general, probabilities
derived from Kolmogorov--Smirnov tests and maximum likelihood fits should be
treated with a degree of caution.  A low probability merely implies that the
model is a poor fit to the data, without specifying what aspect of the model
is different from the data.  Parametric fits can yield spuriously low
probabilities if the functional {\it form\/} of the model is not a good
approximation to the data.  A comparison between the shapes of curves~2 and 3
in Figure~7 and that of the corrected MAPEL estimates of M15's density
profile (Fig.~11) shows that the functional form given in Eq.~(4) does
represent the data well.

Merritt \& Tremblay (1994) derive confidence intervals for the surface
density by generating a large sample of ``bootstrap'' samples from
the MAPEL estimate (cf.~Scott 1992; Taylor 1989).  We have applied this
method to stars with $V<19.0$ within $5''$ of the center of M15.  Figure~12
shows the non-parametric projected density profile corrected for the effects
of incompleteness and photometric error (bold solid line), along with 90\%
(bold dashed line) and 98\% (bold dotted line) confidence bands.  The open
circles show the density profile relative to the alternate center, $0\farcs3$
to the north-west of the centroid (at the location of the concentration of
faint stars); this profile is indistinguishable from the profile centered on
the centroid of the stellar distribution.  An upper
limit to the core radius may be estimated by considering the lowest allowable
central density, ${\rm log}[\sigma(0)]\sim1.2$.  The largest radius at which
the density is consistent with being half the central value, or at which
${\rm log}(\sigma)\sim0.9$, is about $2''$.  This is consistent with the
95\% upper limit on $r_{\rm core}$ derived from maximum likelihood fitting
(Sec.~{\it{5.3}}).

We have used Merritt \& Tremblay's MAPEL method to derive a non-parametric
estimate of the corrected {\it spatial\/} density of stars $\rho(r)$ as a
function of physical radius (in pc).  This is shown in Figure~13 (bold solid
curve) for $V<19$ stars within 0.6~pc ($r<10''$) of the center of M15.  As
before, data within 0.017~pc ($r<0\farcs3$) are excluded from the analysis,
and a smoothing parameter $\lambda=5\times10^{-5}$ is used.  The bold dashed
and dotted lines indicate 90\% and 98\% confidence limits; these limits are
slightly less stringent than those on the projected density profile
(Fig.~12).  It is remarkable that the spatial density of M15 at a radius of
0.02~pc approaches $10^5$~stars~pc$^{-3}$ in post-main-sequence stars alone.
Note, the spatial density profile is more relevant than the projected one in
constraining models.

\vfill\eject
\medskip
\centerline {\it 5.5 Ellipticity of the Isodensity Contours}
\smallskip

The standard picture of globular clusters being non-rotating systems is
beginning to change---recent kinematical studies have demonstrated that
several clusters exhibit a small, but significant amount of rotation in their
inner parts.  Gebhardt \etal\ (1995) have used an imaging Fabry-Perot
spectrophotometer to map the velocity profile of the integrated light in M15
as a function of position and determine a rotation amplitude of
$v_c=2$~km~s$^{-1}$ within $r<10''$ of the cluster center.  The position
angle of the major axis of the velocity field (perpendicular to the axis of
rotation) is roughly $+40^\circ\pm20^\circ$ relative to north.  They find
that the velocity field derived from individual stellar velocities inside
$r<30''$ has a comparable rotation amplitude and orientation.  In M15's
central $5''$, Gebhardt \etal's Fabry-Perot integrated light measurements and
Peterson \etal\ (1989) long-slit spectrum indicate a rotation amplitude as
high as 5~km~s$^{-1}$.  Of these, the integrated light data within $r<10''$
are the least affected by sampling errors.  The line-of-sight stellar
velocity dispersion is $\sigma_v\sim12$~km~s$^{-1}$ in the inner $r<10''$
region of M15 (Dubath \& Meylan 1994; Gebhardt \etal\ 1994), and this implies
a ratio of $v_c/\sigma_v=0.17$.

For an isotropic oblate rotator, the virial theorem predicts that the
isodensity contours should have an ellipticity $e\sim(v_c/\sigma_v)^2$
(Binney 1978), which corresponds to $e\sim3\%$ for the $r<10''$ region of
M15.  Gebhardt \etal\ (1995) have applied an adaptive kernel smoothing
technique to stars detected in pre-refurbishment {\it HST\/} images of M15
(Paper~II) in order to map the isopleths of the stellar density (see also
Merritt \& Tremblay 1994).  The authors remark that the isopleths in the
$r<10''$ region appear to have too high an ellipticity compared to the value
of 3\% expected from the observed rotation.  However, the extended nature of
the PSF in the aberrated images ($r_{\rm PSF}\sim2\farcs5$) implies that any
estimate of the local stellar surface density derived from these images is
difficult to interpret.  A bright star tends to produce a ``hole'' in the
distribution since faint stars ($V\sim19$) are undetectable in its vicinity,
yet faint stars dominate the sample in regions away from bright stars.
The isopleths measured by Gebhardt \etal\ may be distorted by a handful of
bright RGB stars (this is analogous, but opposite in sign, to the effect
these stars have on the overall light distribution).  The PSF in WFPC2 images
has practically no extended structure---thus, these images yield a more
uniform sample down to a fainter stellar detection threshold than aberrated
{\it HST\/} images.

We have used the WFPC2 $V<20.0$ sample of stars to determine the ellipticity
$e$ of the surface density distribution within $r\la15''$ of the center of
M15.  An elliptical model distribution, with a radial density profile of the
form given by Eq.~(4), is fit to the data as described in Sec.~{\it{5.3}},
except that the radial distance $r$ is replaced by the elliptical radial
distance parameter:
$$~~~~~{\tilde r} ~=~ [(x/q)^2\, + \,(yq)^2]^{1/2} ~~~~~~, \eqno(7{\rm a})$$
where
$$~~~~~q ~=~ (1 \, - \, e)^{1/2} ~~~~~~. \eqno(7{\rm b})$$
The $(x,~y)$ axes define the minor and major axes of the model, respectively.
A grid of ellipticities $e$ and position angles are fit to the data, while
$r_{\rm core}$ and $\alpha$ are held fixed at the optimal values derived from
the azimuthally symmetric fit.  Beyond $r=5\farcs8$, the radial density
profile is assumed to be a power law with index $\alpha_2=-1.3$ (Fig.~8).
The models are multiplied by the analytical approximation to the star count
correction factor $C(r)$ given in Eq.~(2) before fitting to the observed
(uncorrected) stellar distribution.

Figure~14 shows the best fit $e$ and position angle ($\times$) along with
75\%, 90\%, and 99\% confidence contours for these parameters based on a
maximum likelihood fit to M15 stars in the radial ranges:
$0\farcs5<r<5\farcs8$ (upper left panel), $5\farcs8\leq{r}<10''$ (upper
right), and $10''<r<15''$ (lower left).  The combined $0\farcs5<r<15''$
sample (lower right panel) contains nearly 5000~stars and has an ellipticity
$e=0.05\pm0.04$ (90\% confidence limits) at a position angle of
$+60^\circ\pm25^\circ$ (measured north through east).  This is consistent
with the ellipticity and orientation expected on the basis of the observed
rotation in the $r<10''$ region of the cluster, to within the errors of both
sets of measurements.  The hypothesis of azimuthal symmetry in the $r<15''$
distribution is rejected at
the 95\% significance level.  It is reassuring that each of the three
independent radial subsamples yields maximum likelihood values of $e$ and
position angle that agree with the values determined from the combined
sample.  The Poisson noise in an individual subsample is too large to permit
independent determination of $e$ as a function of radius.

\bigskip
\centerline {\smcap 6. discussion: the central cusp}
\medskip

The $0\farcs1$ resolution of WFPC2 images makes it possible to study the
crowded inner few arcseconds of M15 at a level of detail that is beyond what
has been achieved with even the highest resolution ground-based images
(cf.~Steton 1994) or with pre-repair {\it HST\/} images (Lauer \etal\ 1991;
Paper~II).  As many as 58~stars with $V<19$ and 82~stars with $V<20$ are
detected in the WFPC2 dataset within the central $r<1''$ circle alone.  This
allows determination of the stellar surface density profile on subarcsecond
scales near the cluster center.

As discussed in the preceding subsections, the radial projected density
profile of M15 rises as a power law towards the center, all the way in to
$r=0\farcs3$.  Given the finite number of stars in any globular cluster, no
matter how dense, there is a minimum radius within which the lack of resolved
stars and/or extreme crowding limit the ability to determine: (1)~the cluster
center, (2)~the observed (uncorrected) central density, and (3)~the star
count correction factor.  For our WFPC2 data of M15, this minimum radius is
about $0\farcs3$.  The density profile of M15 shows no signs of levelling off
in the radial range over which it can be reliably measured (see Figs.~6 and
11).  Poisson errors in the star counts, however, make it impossible to rule
out the presence of a small core: profiles with $r_{\rm core}<1\farcs8$ are
consistent with the data at the $\ga10$\% level (Sec.~{\it{5.3}}).  The data
do not {\it require\/} a finite $r_{\rm core}$; non-parametric estimates of
the density profile are best approximated by profiles with $r_{\rm
core}<1''$.  These limits on $r_{\rm core}$ are likely to be biased high due
to errors in the determination of the cluster centroid.  The half-light (and
presumably half-mass) radius $r_{\rm h}$ of M15 is $61''$ (Djorgovski 1993),
so a core radius of $1''$ (0.06~pc) corresponds to $0.016\>r_{\rm h}$.
Energy injection by close binary stars or the bounce of a core-collapsed
cluster is expected to produce a core whose radius is a few percent of
$r_{\rm h}$ (Goodman 1989).  Grabhorn \etal's (1992) Fokker-Planck simulation
of M15 suggests that the core radius is larger than $1''$ 95\% of the time
and larger than $1\farcs5$ about 75\% of the time.  This is roughly
consistent with the limits on $r_{\rm core}$ imposed by the observations.

The shape of M15's star count profile, $r^\alpha$ with $\alpha=-0.82\pm0.12$,
is remarkably similar to the $\alpha=-0.75$ steady-state solution calculated
by Bahcall \& Wolf (1976, 1977) for the radial distribution of equal mass
stars around a point mass.  The radial extent over which a point mass affects
the orbits of stars (roughly twice the ``gravitational radius'' $r_{\rm g}$
defined in Paper~I) is proportional to its mass.  The density profile in M15
is well approximated by an $\alpha=-0.8$ power law out to $r=6''$, beyond
which it steepens to $\alpha_2=-1.3$ (Fig.~8), a value significantly
different from the Bahcall \& Wolf solution for a compact object.  If we
assume that the power law cusp in M15 is the result of a central black hole,
the mass of the black hole $M_{\rm BH}$ must be about $10^4\,M_\odot$ in
order to affect the stellar distribution out to $r\sim6''$ (see Fig.~10 of
Paper~II).  In fact, this represents an upper limit to the black hole mass
since larger values of $M_{\rm BH}$ would cause the profile slope at $r>6''$
to be shallower than the measured value.

A more restrictive upper limit on $M_{\rm BH}$ comes from measurements of the
line-of-sight velocity dispersion $\sigma_v$ near the center of M15.
Peterson \etal\ (1989) found $\sigma_v\sim25$~km~s$^{-1}$ in the inner few
arcseconds of the cluster; this constrains $M_{\rm BH}$ to be less than
about $3\times10^3\,M_\odot$.  The integrated stellar absorption line profile
near the center of M15, however, is likely to be dominated by a few bright
stars and this makes the determination of $\sigma_v$ uncertain (Zaggia
\etal\ 1992, 1993; Dubath \etal\ 1994).  Recent long-slit spectroscopy of the
central $5''\times8''$ region of M15 by Dubath \& Meylan (1994) suggests that
the velocity dispersion at a radius of $3''$ is about $12\pm3$~km~s$^{-1}$.
This is consistent with the 90\% upper limit of $\sigma_v<16$~km~s$^{-1}$ at
the same radius derived by Gebhardt \etal\ (1994) with their Fabry-Perot
data.  If the {\it central\/} value of $\sigma_v$ is as low as these recent
studies indicate, the upper limit on the central black hole mass is $M_{\rm
BH}\la10^3\,M_\odot$.  Such a black hole would only affect the stellar
distribution out to $r\sim1''$.  The underlying distribution (i.e., that
which would exist in the absence of a black hole) must then be a power law in
the range $1''\la{r}\la6''$ in order to match the observations.  The fact
that M15's density profile is smooth through the $r=1''$ region, extending
all the way out to $r=6''$ without significant departures from an
$\alpha=-0.8$ power law, makes the case for a $M_{\rm BH}<10^3\,M_\odot$
central black hole less compelling.

The measured slope of the surface density profile in M15 is also well within
the range of slopes expected during and after the process of core collapse:
$-0.5\la\alpha\la-1.5$ (Grabhorn \etal\ 1992; Heggie 1985).  An important
discriminant between clusters with a central compact mass and those without
is the central velocity dispersion of the stars.  Since the nature of core
collapse is such that it is a ``cooling'' process, it does not result in a
prominent increase in $\sigma_v$ towards smaller radii.  By contrast, a
central massive black hole causes the dispersion to increase in Keplerian
fashion with decreasing radius, $\sigma_v\propto{r}^{-0.5}$ (Bahcall \& Wolf
1976).  The high angular resolution provided by post-repair {\it HST\/}
optics should facilitate the measurement of M15's central velocity dispersion
and thereby resolve the ambiguity.

The ground-based kinematic measurements of Gebhardt \etal\ (1994) and Dubath
\etal\ (1994) were obtained in about $1''$ seeing.  Even if M15 contains a
cusp of stars with a high velocity dispersion but with a radial extent
smaller than about $0\farcs5$, the light of nearby bright RGB stars would
dominate the absorption line profile causing the cusp to be masked in the
ground-based datasets.  There is an unusual object, AC~214, located within
$r=0\farcs5$ ($1.5\sigma$ error circle) of the cluster centroid, that was
resolved into at least three stars in pre-refurbishment {\it HST\/} images
(Paper~II).  WFPC2 data confirm that the three stars each have a brightness
of $V\sim15.5$ and are separated by only $0\farcs1$ ($10^3\,$AU in
projection).  If the velocity differences between members of this triplet are
found to be high ($\Delta{v}>40$~km~s$^{-1}$), it would provide strong
evidence for a $\ga10^3\,M_\odot$ compact mass in their midst.  There is also
a small clump of fainter stars with a size of $\sim0\farcs2$ located
$0\farcs3$ to the north-west of the cluster centroid (Fig.~2); it is possible
that this stellar agglomeration marks the exact location of M15's density
cusp.

Independent evidence in favor of a mass concentration near the center of M15
comes from accurate timing measurements of two millisecond pulsars.  While
most millisecond pulsars have a positive intrinsic period derivative $\dot P$
(`spin down'), the fact that these two have a negative $\dot P$ suggests that
the pulsars are being accelerated towards the cluster center at a high rate.
The observed period derivative provides a lower limit on the mass surface
density within the central $1\farcs1$ of M15 of
$\sim4\times10^5\,M_\odot\,{\rm{pc}^{-2}}$ (Phinney 1993), which corresponds
to $M>4.5\times10^3\,M_\odot$ within this region.

There are 82~stars with $V<20$ detected within $r<1''$ of the center of M15
in the WFPC2 images, and our simulation predicts an incompleteness correction
factor $1/C=2.5$ (see Fig.~5).  A typical post-main-sequence or turnoff star
($V\la20$) has a mass of $0.75\,M_\odot$ (Bergbusch \& VandenBerg 1992).  The
$V<20$ stars account for about $150\,M_\odot$.  The amount of mass in stars
fainter than $V=20$ is not well constrained by the data.  The diffuse light
of such stars is much fainter than the light of the giants; the process of
subtracting the RGB light and measuring the residual light is very uncertain.
If the stellar mass function in the center of M15 rises as steeply towards
lower masses as it does in the outer parts of the cluster, main sequence
stars would be expected to contribute more than ten times as much mass as
is found in $V<20$ stars.  For example, the Bergbusch \& VandenBerg 15~Gyr
metal poor isochrone for a Salpeter mass function [$N(m)\propto{m}^{-(1+x)}$,
with $x=1.35$], normalized to contain $150\,M_\odot$ in stars brighter than
$V=20$ ($M_V<4.8$), predicts nearly $4\times10^3\,M_\odot$ down to a lower
mass limit of $0.2\,M_\odot$.  Their model LF is a very good approximation to
the shape of the LF of post-main-sequence stars ($V<19$) beyond $r>1'$ in M15
(where the data are complete down to $V=21$; see Fig.~3), but overpredicts
the number of $V=21$ stars by $\sim50$\%.  More importantly, mass segregation
is expected to cause the mass function to be shallower in the center than it
is in the outskirts.  Phinney (1993) argues that most of the
$4.5\times10^3\,M_\odot$ needed to produce the pulsar acceleration cannot be
in the form of very low mass stars and must be primarily in the form of dark
remnants (white dwarfs and neutron stars).  Concentrating most of the mass in
one compact central object of mass $\ga10^3\msol$ would also explain the
pulsar acceleration data.

\bigskip
\centerline {\smcap 7. summary}
\medskip

\item{\bf 1.}{We have obtained F336W, F439W, and F555W (approximately $UBV$)
photometry of more than $3\times10^4$~stars in the central 5~arcmin$^2$
($r<2'$) of the dense globular cluster M15 (NGC~7078), using {\it Hubble
Space Telescope\/} Wide Field and Planetary Camera~2 images.  A new
analysis technique has been applied to the data---one that combines point
spread function fitting and aperture photometry.  This paper describes
measurement of the surface density distribution of the cluster based on $V$
magnitude-limited samples of stars.}

\smallskip
\item{\bf 2.}{A realistic simulated image has been constructed.  This image
has been analysed in the same manner as the M15 WFPC2 data.  Photometric
accuracy in the inner $15''$ of M15 is estimated to be $1\sigma\la0.1$~mag
for stars with $V<20$, which is 1~mag below the main sequence turnoff.  The
simulation is used to quantify the effect of incompleteness and photometric
error on magnitude-limited star counts so that the M15 samples may be
corrected.  Star identifications are nearly complete for $V<19$ even in the
dense cluster center.}

\smallskip
\item{\bf 3.}{The number-weighted centroid of the stellar distribution,
calculated from samples of various limiting magnitudes and radii, is
used to define the center of M15.  The $1\sigma$ positional error in the
cluster center is about $0\farcs3$.  The triple star AC~214 lies about
$0\farcs5$ ($1.5\sigma$) east of the cluster centroid, while a dense
concentration of faint stars is located $0\farcs3$ north-west of the
centroid.}

\smallskip
\item{\bf 4.}{We have presented three complementary approaches to measuring
the surface density distribution in M15: binned star counts, parametric fits,
and non-parametric estimates.  The projected density profile of
post-main-sequence stars,
corrected for incompleteness and photometric error, is well approximated by a
power law, $r^\alpha$, with $\alpha=-0.82\pm0.12$ (95\% limits) in the range
$0\farcs3<r<6''$.  The density appears to rise smoothly towards the center
over this region of the cluster with no suggestion of levelling off.  More
than 80~stars are detected within the central $r=1''$ circle in M15; this
permits reliable determination of the density profile in the range
$0\farcs3<r<1''$.  Uncertainties in the cluster centroid position and in the
star count correction factor, along with Poisson error in the counts,
restrict the analysis of M15's surface density profile to $r>0\farcs3$.
The 95\% upper limit on the core radius is $2''$.}

\smallskip
\item{\bf 5.}{The profile slope in the inner $6''$ of M15 is very similar to
that expected for the stellar distribution around a black hole with a mass of
a few times $10^3\,M_\odot$.  The observed density profile is also consistent
with core-collapse models.  High angular resolution kinematical measurements
are required to distinguish between these two possibilities.}

\smallskip
\item{\bf 6.}{The stellar distribution within $r\la15''$ of the center of M15
shows a 5\% ellipticity, and departs from circular symmetry at the 95\%
level.  The shape and orientation of the isodensity contours are consistent
with the amplitude and direction of rotation observed by Gebhardt \etal\
(1995) in the inner parts of this cluster.}

\bigskip\bigskip

We would like to thank D.~Merritt for providing MAPEL routines, P.~Stetson
for providing {\smcap daophot ii} software, and M.~Dickinson for assistance
in the preparation of the color image.  We
acknowledge useful discussions with M.~Bolte, K.~Gebhardt, P.~Hut, T.~Lauer,
D.~Merritt, and H.-W.~Rix.  P.G.\ would like to thank the Institute for
Advanced Study for its generous hospitality.  This work was supported in part
by NASA through grant number NAG5-1618 and by STScI grant GO-5324.01-93A.

\bigskip\medskip
\centerline {\smcap references}
\smallskip

\reference Auri\`ere, M., \& Cordoni, J.-P. 1981, A\&AS, 46, 347

\reference Bahcall, J.~N., Bahcall, N.~A., \& Weistrop, D. 1975, Astrophys.\
Lett., 16, 159

\reference Bahcall, J.~N., \& Ostriker, J.~P. 1975, Nature, 256, 23

\reference Bahcall, J.~N., \& Wolf, R.~A. 1976, ApJ, 209, 214

\reference Bahcall, J.~N., \& Wolf, R.~A. 1977, ApJ, 216, 883

\reference Benz, W., \& Hills, J.~G. 1987, ApJ, 323, 614

\reference Bergbusch, P.~A., \& VandenBerg, D.~A. 1992, ApJS, 81, 163

\reference Binney, J.~J. 1978, MNRAS, 183, 501

\reference Bolte, M. 1989, ApJ, 341, 168

\reference Burrows, C. J. (Ed.) 1994, Wide Field and Planetary Camera~2
Instrument Handbook, Version 2.0, Space Telescope Science Institute
publication

\reference DeMarchi, G., \& Paresce, F. 1994, ApJ, 422, 597

\reference Djorgovski, S.~G. 1993, in Structure and Dynamics of Globular
Clusters, edited by S.~G.~Djorgovski and G.~Meylan (ASP Conf.\ Series,
No.~50), p.~373

\reference Djorgovski, S.~G., \& King, I.~R. 1984, ApJ, 277, L49

\reference Dubath, P., Meylan, G., \& Mayor, M. 1994, ApJ, 426, 192

\reference Dubath, P., \& Meylan, G. 1994, A\&A, 290, 104

\reference Efron, B. 1982, The Jacknife, the Bootstrap, and Other Resampling
Plans (Philadelphia: SIAM)

\reference Fahlman, G.~G., Richer, H.~B., \& VandenBerg D.~A. 1985, ApJS, 58,
225

\reference Ferraro, F.~R., \& Paresce, F. 1993, AJ, 106, 154

\reference Gao, B., Goodman, J., Cohn, H.~N., \& Murphy, B.~W. 1991, ApJ,
370, 567

\reference Gebhardt, K., Pryor, C., Williams, T.~B., \& Hesser, J.~E. 1995,
AJ, 110, 1699

\reference Gebhardt, K., Pryor, C., Williams, T.~B., \& Hesser, J.~E. 1994,
AJ, 107, 2067

\reference Goodman, J. 1989, in Dynamics of Dense Stellar Systems, edited by
D.~Merritt (Cambridge University Press, New York), p.~183

\reference Goodman, J., \& Hut, P. 1989, Nature, 339, 40

\reference Grabhorn, R.~P., Cohn, H.~N., Lugger, P.~M., \& Murphy, B.~W.
1992, ApJ, 392, 86

\reference Guhathakurta, P., Yanny, B., Bahcall, J. N., \& Schneider, D.~P.
1994, AJ, 108, 1786

\reference Guhathakurta, P., Yanny, B., Schneider, D.~P., \& Bahcall, J.~N.
1992, AJ, 104, 1790 (Paper~I)

\reference Heggie, D.~C. 1975, MNRAS, 173, 729

\reference Heggie, D.~C. 1985, in Dynamics of Star Clusters, Proc.\ IAU
Symposium No.~113, edited by J.~Goodman and P.~Hut (Reidel, Dordrecht),
p.~139

\reference Heggie, D.~C., \& Aarseth, S.~J. 1992, MNRAS, 257, 513

\reference Hut, P., \etal\ 1992a, PASP, 104, 681

\reference Hut, P., McMillan, S., \& Romani, R.~W. 1992b, ApJ, 389, 527

\reference King, I.~R. 1975, in Dynamics of Stellar Systems, Proc.\ IAU
Symposium No.~69, edited by A.~Hayli (Reidel, Dordrecht), p.~99

\reference Kulkarni, S.~R., Goss, W.~M., Wolszczan, A., \& Middleditch, J.~M.
1990, ApJ, 363, L5

\reference Lauer, T.~R., \etal\ 1991, ApJ, 369, L45

\reference Leonard, P.~J. 1989, AJ, 98, 217

\reference Leonard, P.~J., \& Fahlman, G.~G. 1991, AJ, 102, 994

\reference Lugger, P.~M., Cohn, H.~N., Grindlay, J.~E., Bailyn, C.~D., \&
Hertz, P. 1987, ApJ, 320, 482

\reference McMillan, S.~L.~W. 1989,
in Dynamics of Dense Stellar Systems, edited by D.~Merritt (Cambridge
University Press, New York), p.~207

\reference McMillan, S.~L.~W., Hut, P., \& Makino, J. 1991, ApJ, 372, 111

\reference Merritt, D., \& Tremblay, B. 1994, AJ, 108, 514

\reference Murphy, B.~W., Cohn, H.~N., \& Hut, P. 1990, MNRAS, 245, 335

\reference Paresce. F., \etal\ 1991, Nature, 352, 297

\reference Peterson, R.~C., Seitzer, P., \& Cudworth, K.~M. 1989, ApJ, 347,
251

\reference Phinney, E.~S. 1993, in Structure and Dynamics of Globular
Clusters, edited by S.~G.~Djorgovski and G.~Meylan (ASP Conf.\ Series,
No.~50), p.~141

\reference Pryor, C., Smith, G.~H., \& McClure, R.~D. 1986, AJ, 92, 1358

\reference Pryor, C., McClure, R.~D., Hesser, J.~E., \& Fletcher, J.~M. 1989,
in Dynamics of Dense Stellar Systems, edited by D.~Merritt (Cambridge
University Press, New York), p.~175

\reference Scott, D.~W. 1992, Multivariate Density Estimation (New York:
Wiley)

\reference Stetson, P.~B.  1987, PASP, 99, 191

\reference Stetson, P.~B. 1992, in Astronomical Data Analysis Software,
edited by D.~M.~Worrall, C.~Biemesderfer, and J.~Barnes (ASP Conf.\ Series,
No.~25), p.~297

\reference Stetson, P.~B. 1994, PASP, 106, 250

\reference Sugimoto, D., \& Bettwieser, E. 1983, MNRAS, 204, 19P

\reference Taylor, C.~C. 1989, Biometrika, 76, 705

\reference Thompson, J.~R., \& Tapia, R.~A. 1990, Nonparametric Function
Estimation, Modeling, and Simulation (Philadelphia: SIAM)

\reference Trauger, J.~T., \etal\ 1994, ApJ, 435, L3

\reference Webbink, R.~F. 1985, in Dynamics of Star Clusters, Proc.\ IAU
Symposium No.~113, edited by J.~Goodman and P.~Hut (Reidel, Dordrecht),
p.~541

\reference Yanny, B., Guhathakurta, P., Bahcall, J.~N., \& Schneider, D.~P.
1994a, AJ, 107, 1745 (Paper~II)

\reference Yanny, B., Guhathakurta, P., Schneider, D.~P., \& Bahcall, J.~N.
1994b, ApJ, 435, L39 (Paper~IV)

\reference Zaggia, S., Capaccioli, M., \& Piotto, G. 1993, A\&A, 278, 415

\reference Zaggia, S., Capaccioli, M., Piotto, G., \& Stiavelli, M. 1992,
A\&A, 258, 302

\vfill\eject
\centerline {\smcap figure captions}
\bigskip

\parskip=0pt

\medskip
\noindent
{\smcap fig.~1.}~~
Greyscale representation of the median of four~$V$ band {\it HST\/} WFPC2
images (each with an exposure time of 8~s) of the central region of~M15.  The
WFPC2 mosaic covers an area of~5~arcmin$^2$ and consists of a
$34''\times34''$ PC1 CCD image and three WF CCD images each $80''$ on a side.
The gaps shown in the figure do not accurately represent the slight rotations
and actual gaps between the fields of view of adjacent CCDs.  The cluster
center is located in the PC1 CCD.  The scale and orientation of the image are
indicated in the upper right.

\medskip\smallskip\smallskip
\noindent
{\smcap fig.~2.}~~
A ``true color'' (24-bit) image of the central $9''\times9''$ region of M15
imaged on the PC1 CCD.  The red, green, and blue intensities are proportional
to the brightnesses in the $V$, $B$, and $U$ bands, respectively.  Bright red
giant branch stars with $B-V\sim1.3$ appear reddish orange while blue
horizontal branch stars with $B-V\sim0$ appear blue.  A green cross marks the
cluster center determined from the number-weighted centroid of stellar
positions.  This image has the same orientation as Figure~1.  The surface
density of stars is clearly seen to increase towards the cluster center all
the way in to radii less than $0\farcs5$.  The X-ray source AC~211 is the
very blue star located roughly $2''$ north of the cluster center.  There is a
concentration of faint stars $0\farcs3$ north-west of the cluster centroid.
The triple star AC~214 is the white object $0\farcs5$ east of the center; its
three components appear blended together in this picture.

\medskip\smallskip\smallskip
\noindent
{\smcap fig.~3.}~~
Luminosity function of stars in bins of increasing radial distance from the
center of M15.  The turnover at faint magnitudes is due to incompleteness.
The limits of the radial bins have been chosen so that they include equal
numbers of stars brighter than $V=19$, the main sequence turnoff point.  The
extrapolated luminosity function used for the simulations in Sec.~3 is
plotted as open circles (arbitrary normalization).  The bump at $V\sim16$ is
caused by horizontal branch stars.

\medskip\smallskip\smallskip
\noindent
{\smcap fig.~4.}~~
Photometric error as a function of $V$ magnitude as determined from our
simulation of the M15 PC1 image.  The top panel shows the difference between
the input (``true'') and output (measured) magnitude for each of
$\ga10^4$~artificial stars.  The distribution is skewed towards positive
values at faint magnitudes---i.e., the measured magnitudes are systematically
too bright---as a result of blending.  Stars are divided into two broad
radial bins, $r<5''$ (middle panel) and $5''<r<20''$ (bottom panel), and are
grouped into 0.5~mag bins in order to compute the mean
photometric bias and $1\sigma$ rms scatter (open circles and error bars).
The median bias is indicated by solid dots and the associated error bars mark
the range that includes $\pm34\%$ of the stars on either side of this median
value.  (A slight horizontal shift has been introduced between the open and
and solid symbols for the sake of clarity.)  The dashed horizontal lines
indicate error levels of $\pm0.10$~mag.  The $1\sigma$ photometric error
beyond $r>5''$ is about 0.03~mag down to $V=17$ and about 0.2~mag at
$V=20.5$, and the magnitude bias is negligible for $V<21$.  In the inner
regions, the photometric error and bias are significantly worse than in the
outer regions for $V>17$, and degrade rapidly beyond $V=20$.

\medskip\smallskip\smallskip
\noindent
{\smcap fig.~5.}~~
Star count correction factors [Eq.~(1)] derived from the simulation for stars
with $V<18.3$ (top), $V<19.0$ (middle), and $V<20.0$ (bottom).  The points
are based on ratios of binned star counts, the error bars indicate Poisson
errors in the number of output stars, and the smooth curves are the
exponential approximations given by Eq.~(2).  At bright limiting magnitudes,
$C(r)$ exceeds unity due to the effects of photometric scatter/bias and a
steeply rising LF (Sec.~{\it{4.3}}), while incompleteness dominates at faint
magnitudes.

\medskip\smallskip\smallskip
\noindent
{\smcap fig.~6.}~~
The surface density of stars in the inner $5''$ of M15 binned into
$0\farcs4$-wide annuli, for two independent samples: $V<18.3$ and
$18.3<V<20.0$.  The dashed lines are based on uncorrected (raw) counts while
the bold open squares have been corrected for the effects of incompleteness
and photometric error.  Poisson errors ($1\sigma$) in the raw counts are
indicated.  The shapes of the corrected density profiles derived from the
two samples are similar, even though their raw count profiles are very
different.

\medskip\smallskip\smallskip
\noindent
{\smcap fig.~7.}~~
Corrected density of star counts versus radius in the central $5''$ of M15
for $V<19.0$ (bold open squares) and $V<20.0$ (crosses).  The latter have
been divided by a factor of~4 (0.6~dex) to match the $V<19.0$ profile.  (For
the sake of clarity, the two sets of data points have also been shifted by a
small amount along the horizontal axis relative to each other.)
For comparison, we plot a power law profile with $\alpha=-0.82$ (curve~1),
and profiles with $r_{\rm core}=1''$ (curve~2) and $2''$ (curve~3) using the
formula in Eq.~(4) with the asymptotic slope set to $-0.8$.  The $r_{\rm
core}=2''$ profile appears to be inconsistent with the M15 data.  It is
important to note that estimates of the density interior to
$r<0\farcs3$ (dotted vertical line) are likely to be biased low due to
uncertainties in the position of the cluster center.  The bold dashed line
shows the corrected $V<19.0$ density profile measured relative to an
alternate cluster center, one that is located $0\farcs3$ north-west of the
centroid.

\medskip\smallskip\smallskip
\noindent
{\smcap fig.~8.}~~
Stellar surface density profile of M15 covering the full radial range
($r<100''$) available on the WFPC2 image.  All $V<19$ stars are plotted with
no correction for the effects of incompleteness/photometric error.  The
uncorrected data are well fit by three power law segments, with indices of
$\alpha=-0.7$, $-1.3$, and $-2.0$ (steepening with increasing radius) and
transition radii of $r_{\rm break}=5\farcs8$ and $30''$.  The solid lines
trace the distribution function used in the construction of the simulated
image.

\medskip\smallskip\smallskip
\noindent
{\smcap fig.~9.}~~
Cumulative radial distribution of $V<19$ stars between $0\farcs3<r<5''$ from
the center of M15 (thin stepped line).  Data interior to $r=0\farcs3$ are
excluded since our estimate of the density there is likely to be biased by
errors in the centroid determination.  The upper panel shows power law model
profiles with $\alpha=-0.70$, $-0.82$, and $-0.95$, while the lower panel
shows profiles with cores of radii $r_{\rm core}=1''$ and $2''$ and an
asymptotic power law index $\alpha=-0.8$ [Eq.~(4)].  All model profiles have
been multiplied by the exponential approximation to the star count correction
factor [Eq.~(2)].

\medskip\smallskip\smallskip
\noindent
{\smcap fig.~10.}~~
Results of maximum likelihood fitting of a two-parameter ($r_{\rm core},
{}~\alpha$) family of model profiles, of the form given by Eq.~(4), to stars
between $0\farcs3<r<5''$ in M15.  The upper and lower panels present $V_{\rm
lim}=19.0$ and 20.0 samples, respectively.  Alternate dashed and solid lines
mark 50\%, 75\%, 90\%, 95\%, and 99\% confidence contours.  There is
significant covariance between the core radius and asymptotic slope.  An
$\alpha=-0.82\pm0.12$ (95\% limits) power law fits the data well.  An $r_{\rm
core}=2''$ profile is ruled out at the 95\% level.

\medskip\smallskip\smallskip
\noindent
{\smcap fig.~11.}~~
Non-parametric estimates of the stellar surface density in M15 obtained by
applying Merrit \& Tremblay's (1994) MAximum PEnalized Likelihood (MAPEL)
method.  Three stellar samples with $V_{\rm lim}=18.3$, 19.0, and 20.0 are
presented.  Dashed lines represent uncorrected profiles, while solid lines
have been corrected by dividing by $C(r)$ from Eq.~(2).  The three corrected
profiles are remarkably similar in shape.  The profiles show no indication of
levelling off at small radii.  Their shapes are well approximated by the
analytical form given in Eq.~(4).

\medskip\smallskip\smallskip
\noindent
{\smcap fig.~12.}~~
Non-parametric MAPEL estimate of the surface density of $V<19.0$ stars in M15
(solid line), corrected for incompleteness/photometric error, with 90\% (bold
dashed line) and 98\% (bold dotted line) confidence bands.  The open circles
show the density profile measured relative to an alternate cluster center,
located $0\farcs3$ north-west of the centroid.  An $r_{\rm core}=2''$ profile
can barely be accommodated within the confidence bands.  The most likely
value of the core radius, however, is $<1''$.

\medskip\smallskip\smallskip
\noindent
{\smcap fig.~13.}~~
The {\it spatial\/} density of post-main-sequence stars in M15 (solid line),
a non-parametric (MAPEL) estimate based on the corrected $V<19.0$ sample,
plotted against the radial distance in parsec.  The bold dashed and dotted
lines indicate 90\% and 98\% confidence limits, respectively.  The data are
unreliable interior to $r=0.02$~pc ($0\farcs3$).  The central density of
post-main-sequence stars in M15 appears to exceed $10^5$~stars~pc$^{-3}$.

\medskip\smallskip\smallskip
\noindent
{\smcap fig.~14.}~~
Results of maximum likelihood fitting to determine the ellipticity $e$ and
position angle of the stellar isodensity contours in M15.  Stars are grouped
into three radial bins: $0\farcs5<r<5\farcs8$ (upper left panel),
$5\farcs8<r<10''$ (upper right), and $10''<r<15''$ (lower left).  The full
sample ($0\farcs5<r<15''$) is shown in the lower right panel.  The `$\times$'
marks the best fit values of the parameters, while the contours indicate
75\%, 90\%, and 99\% confidence boundaries.  The number of stars in the set
is indicated.  The combined sample has an ellipticity $e=0.05$ at a position
angle of $+60^\circ$ east of north.  The probability of the stellar
distribution in this sample being azimuthally symmetric is only 5\%.  While
the radial subsamples contain too few stars to permit accurate determination
of $e$ and the position angle, the maximum likelihood solution for each is
consistent with that of the combined sample.

\vfill\eject
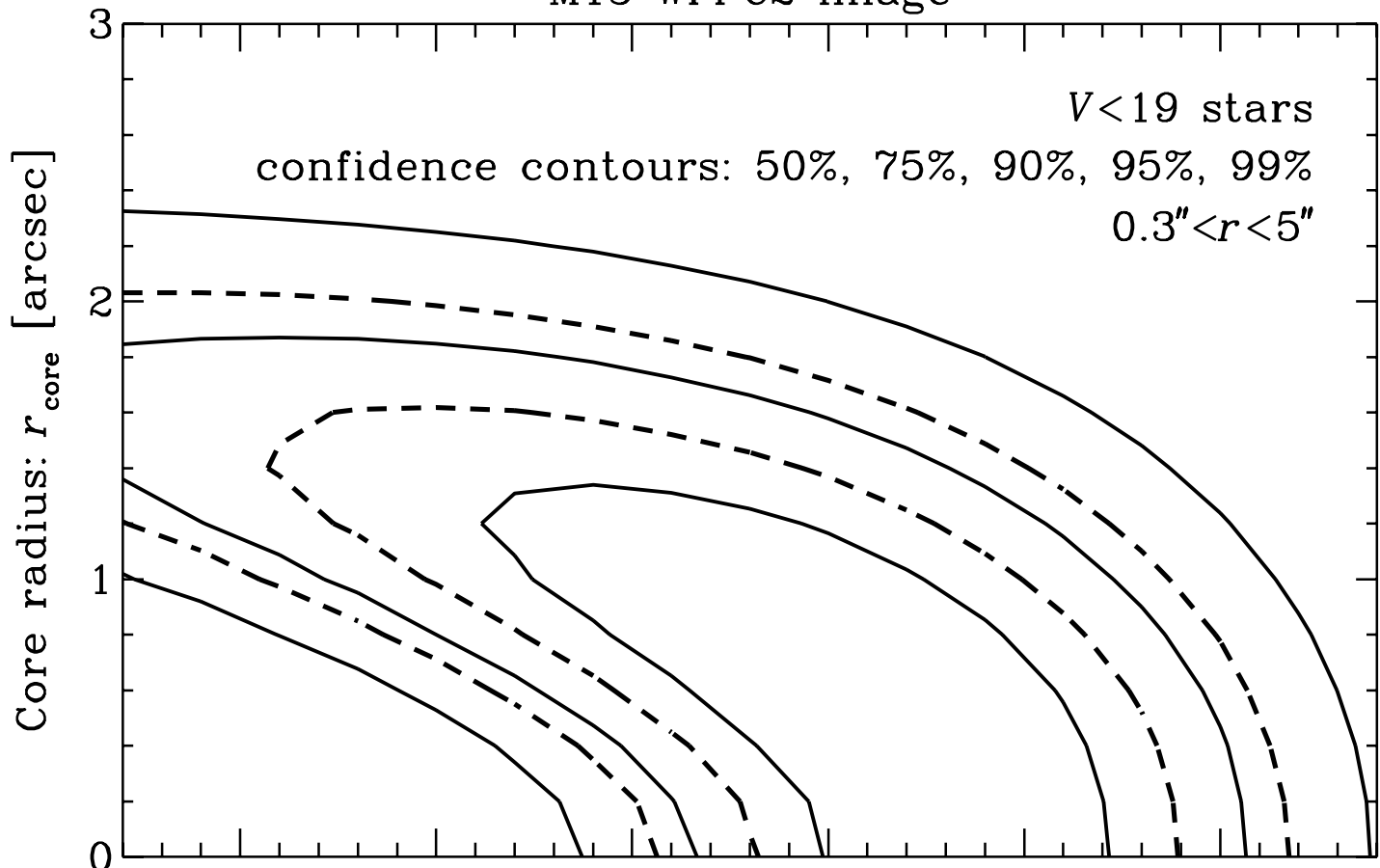
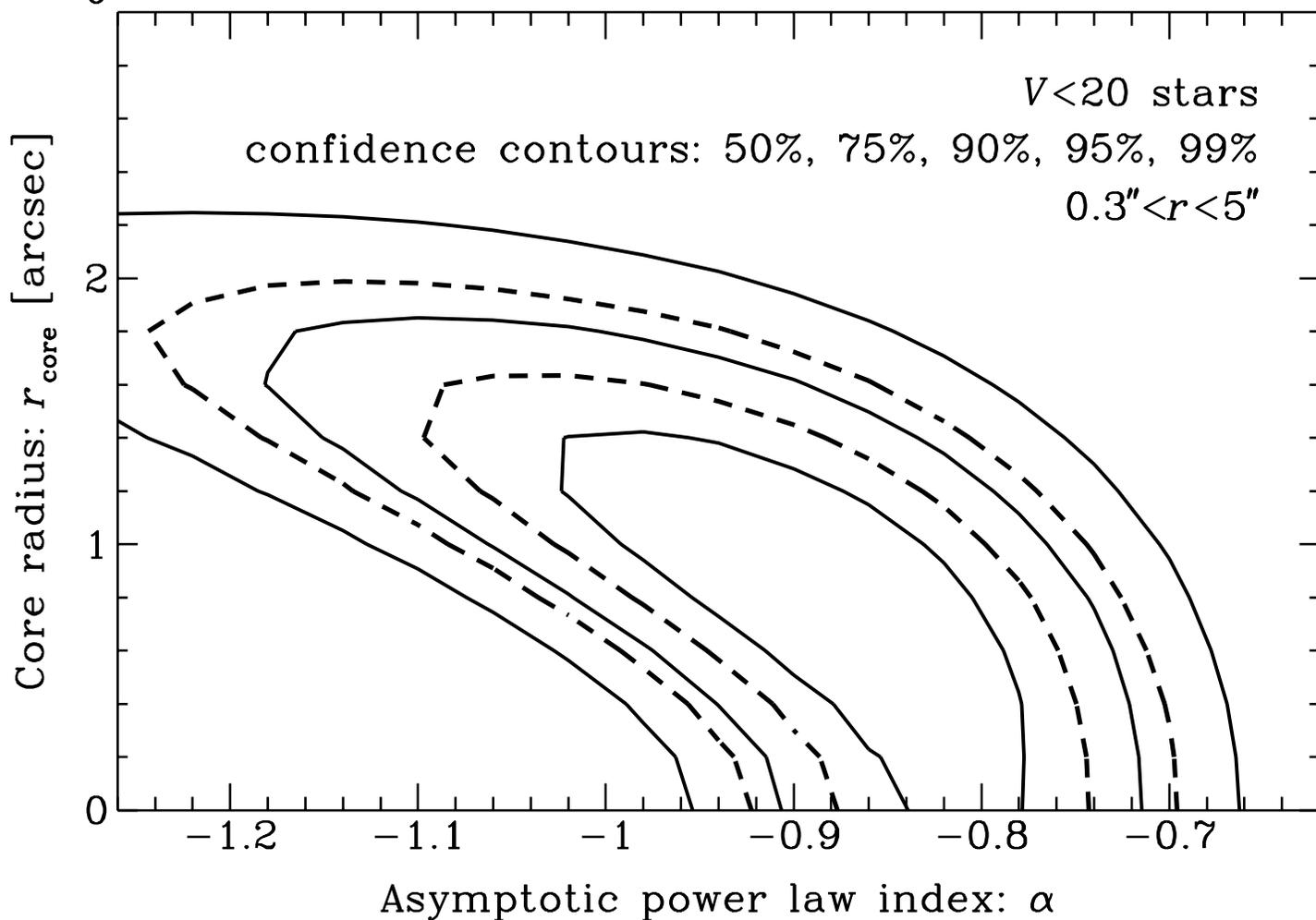
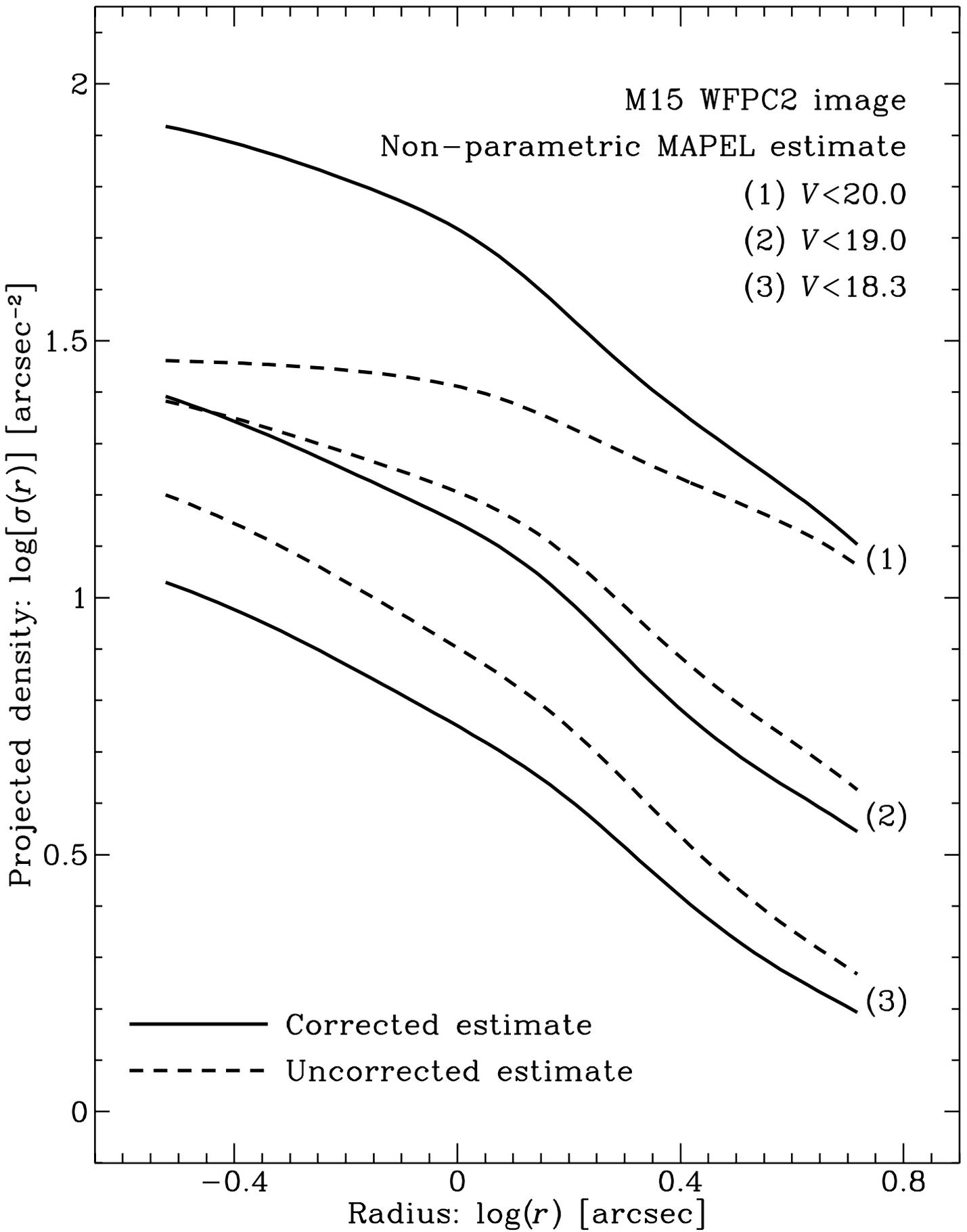
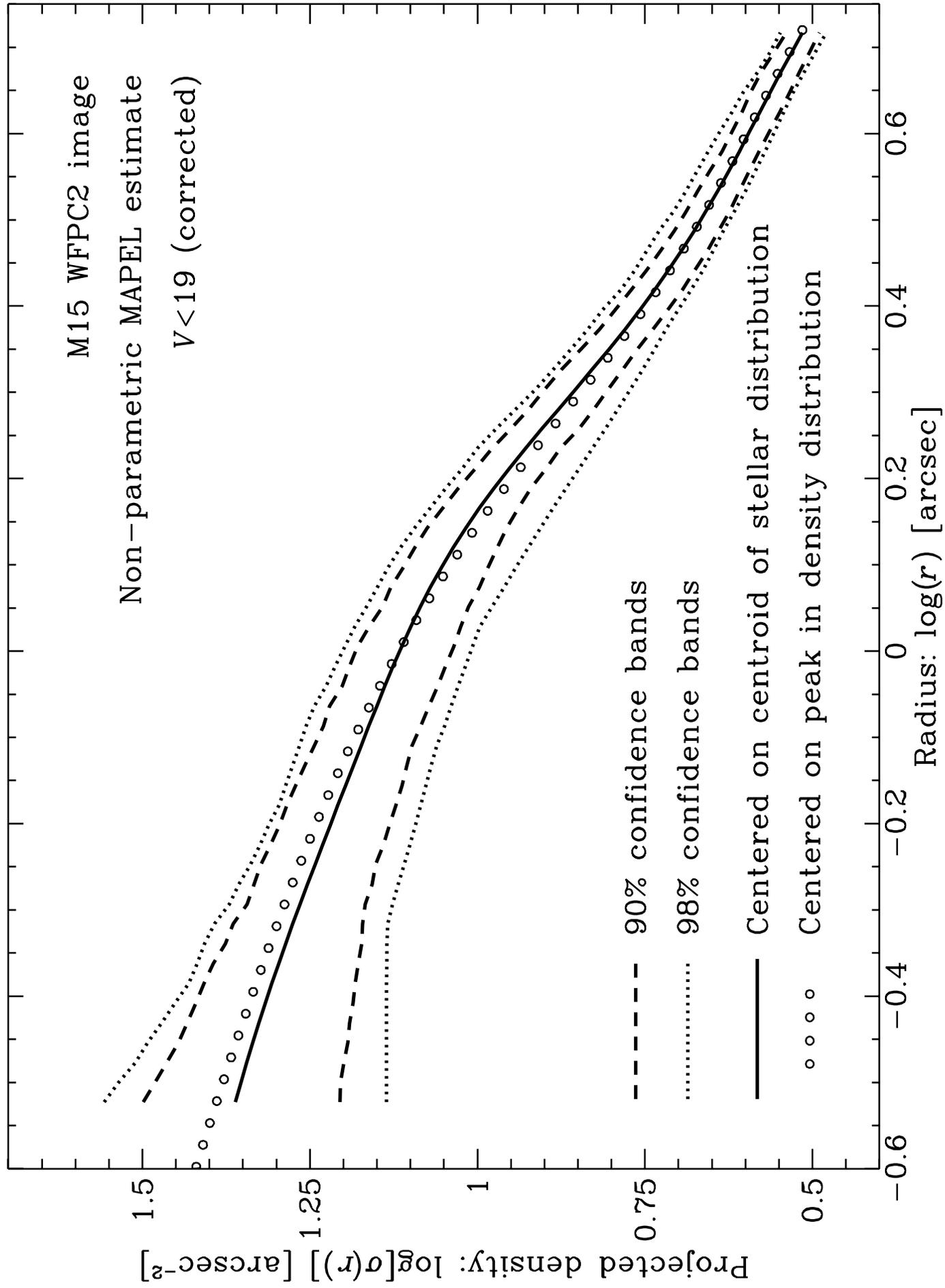
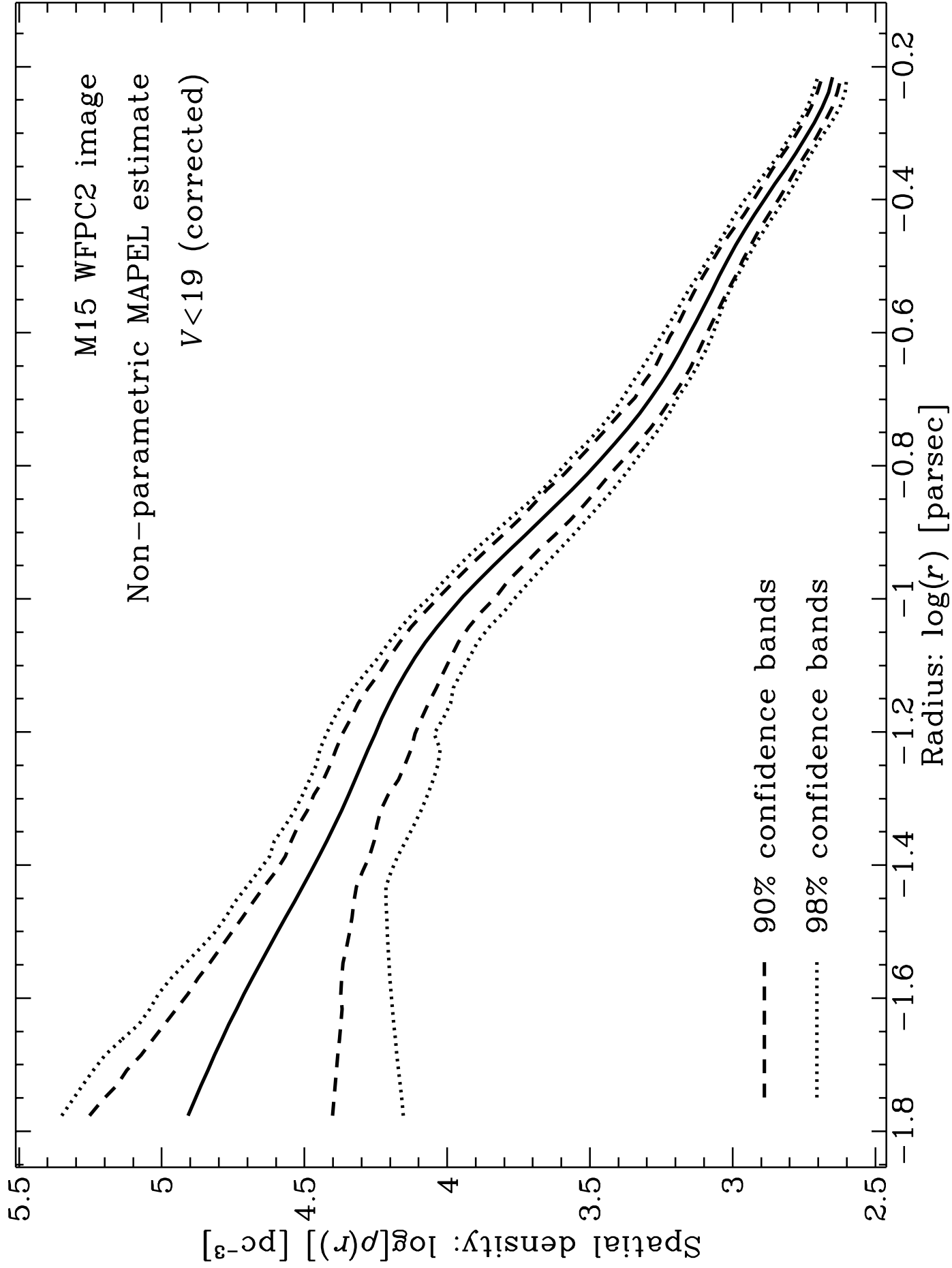
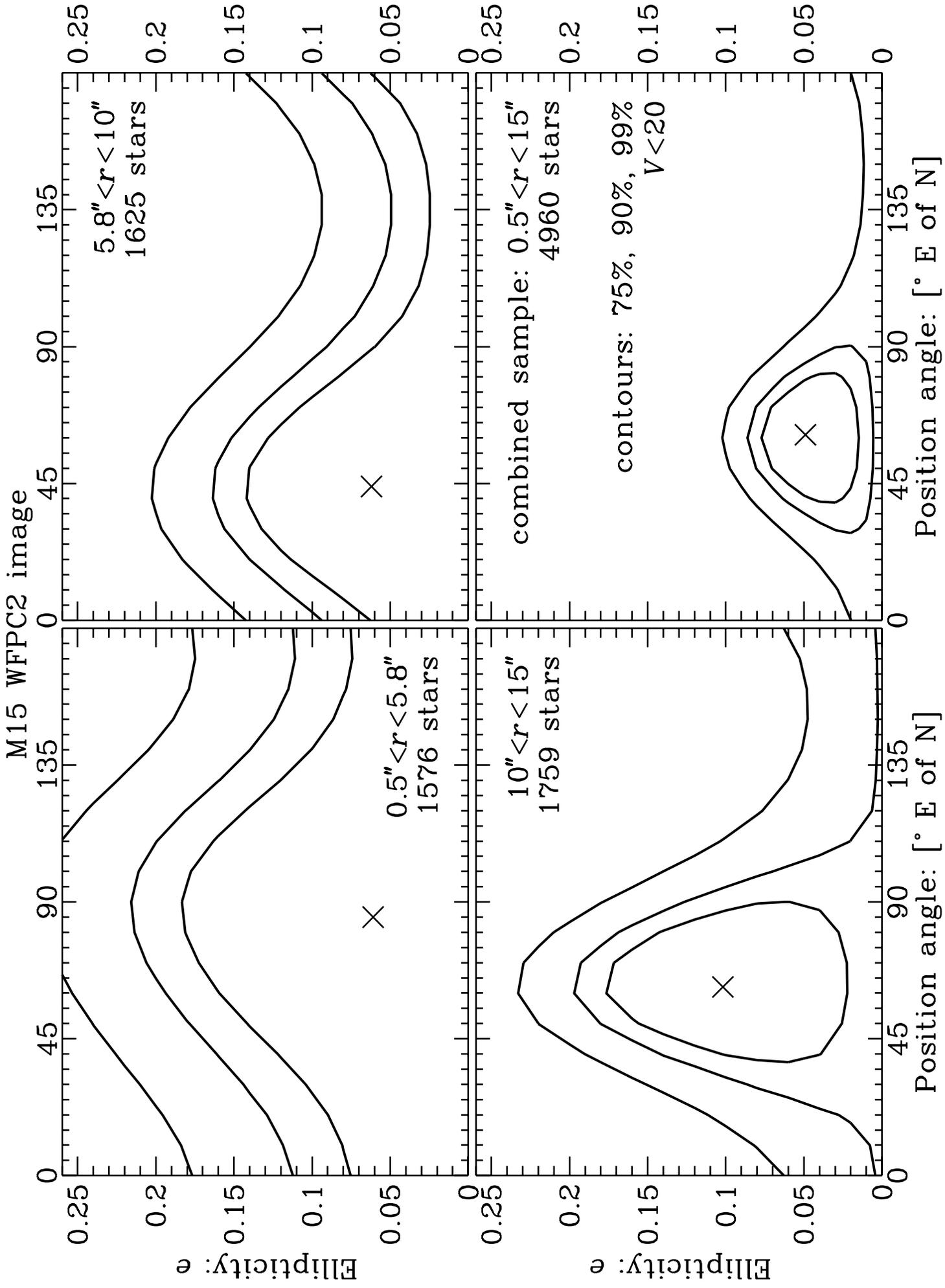
\end